\documentclass[twocolumn,showpacs,preprintnumbers,amsmath,amssymb]{revtex4}
\usepackage{graphicx,epsfig}
\usepackage{rotate}
\usepackage{color}
\usepackage{amsmath,amssymb}

\begin{document}

\title{Synchronization in networks of mobile oscillators}

\author{Naoya Fujiwara}
%\email{fujiwara@pik-potsdam.de}
\affiliation{Potsdam Institute for Climate Impact Research (PIK), 14473 Potsdam, Germany}
\author{J\"{u}rgen Kurths}
%\email{Juergen.Kurths@pik-potsdam.de}
\affiliation{Potsdam Institute for Climate Impact Research (PIK), 14473 Potsdam, Germany}
\author{Albert D\'{\i}az-Guilera}
%\email{albert.diaz@ub.edu}
%\homepage{http://complex.ffn.ub.es/albert}
\affiliation{Departament de F\'{\i}sica Fonamental, Universitat de Barcelona, 08028 Barcelona, Spain}
\affiliation{Potsdam Institute for Climate Impact Research (PIK), 14473 Potsdam, Germany
}

\begin{abstract}
We present a  model of synchronization in networks of
 autonomous agents where the topology changes due to agents motion.
{We introduce two time scales, one for the topological change
 and another one for  local synchronization.
If the former scale is much shorter,} 
%If the topology changes fast,
an approximation 
that averages out the effect of motion is
available.
{Here we show, however,} that
the time required for synchronization achievement
increases with respect to that approximation in the opposite case.
We find that this effect is more important close to the continuum
 percolation transition point.
The simulation results are confirmed by means of
 spectral analysis of the time dependent Laplacian matrix.
Our results show that 
the trade-off between 
%agent speed and interaction frequency,
these two time scales, which have opposite effects
on synchronization,
 should be taken into account for the design
of mobile device networks.
\end{abstract}
\maketitle

After an initial period of characterizing complex networks in terms of local and global statistical
properties (e.g., \cite{ba02}), attention turned to the dynamics of their interacting
units \cite{blmch06}. 
A widely studied example of such behavior is synchronization of coupled oscillators
arranged into complex networks. 
The interplay between topology and dynamics is
crucial  for synchronization achievement (see Ref. \cite{adkmz08} and
references therein).
  In most studies of such systems the network has a fixed structure, 
but there are also many interesting scenarios where the topology changes
over time in various fields such as power transmission system \cite{scl00},
consensus problem \cite{ofm07}, mobile communication \cite{onnela07}, and functional brain
networks \cite{vmdc08}.

%Evolving topologies can be found in 
%\textcolor{red}{many real networks \cite{onnela07,vmdc08,scl00,ofm07}}
%in social networks where links are active only within
%certain time windows \cite{onnela07}, in functional brain networks where the various areas
%are not all connected simultaneously \cite{vmdc08}, in technological networks where
%frequent failures and disruptions mask the flow of information \cite{scl00}, in consensus
%problems \cite{ofm07}, and in several other contexts. 
Within the general framework of time dependent or evolving
networks, we can
identify the particular case of networks whose nodes represent physical agents that move around but interact with each other only when they are close enough. Examples include the
coordinated motions of robots \cite{bffr06}, vehicles \cite{tjp03}, 
and groups of animals \cite{bschdms06}, 
in which cooperative dynamics emerge. 
Especially, there are many examples where synchronization plays a crucial
role: chemotaxis \cite{t07}, mobile ad hoc networks
\cite{r01}, and wireless sensor
networks \cite{sy04}.
Despite the
importance of this topic, prior research on synchronization 
in time-dependent networks of populations of agents has concentrated 
so far on two special cases: i) where the network topology changes fast
{\cite{bbh04b,frasca08,psbs06,sbr06}},
 and ii) where the population is dense and arranged
in a ring \cite{pnm10}.
In the former case, the fast-switching
approximation (FSA) which averages out the effect of agent motion is
commonly used.
However, 
%since synchronization is achieved due to the interplay between 
%instantaneous network topology, agent motion,
%and interaction rules, 
for better understanding of synchronization of mobile
agents and  
design of an efficient network,
it is very important to clarify
when and how FSA fails.

In order to study this point,
this letter proposes a general framework in which agents
perform random walks in a two-dimensional (2D) plane.
We consider that each agent possesses a mobile wireless device
whose state is characterized by a phase variable,
%We assume an interaction %term 
%that makes 
and the phases approach one another
%, but this interaction only 
through %takes place only 
the interaction between agents within a certain spatial range.
This model is well suited for  communication problems with
short-range wireless devices.
%, e.g.,
% for
%example, the abstract mobile devices could easily be replaced 
%{mobile ad hoc network, wireless sensor network} \cite{moays09}. 
%\textcolor{blue}{A reference for mobile ad-hoc and another for wireless sensor?}
%
%To generalize the model, we introduce two different time scales
%for the motion of the agents and the frequency of phase updates.
%We will observe that the dominant time scale changes
%depending on the parameters of the system, and that a number of different
%mechanisms can lead to complete synchronization.
In this letter we show a general mechanism of failure of FSA
when the time scale of local synchronization
is
shorter than the time scale of the topology change due to the agent
motion.
Since we need longer synchronization time due to this failure,
it is an important factor we should take into account for constructing
an efficient mobile network.

Our model consists of $N$ agents moving in a 2D space (size
$L\times L$) with periodic boundary conditions.
Each agent moves with velocity $v$. 
%during 
%time intervals of length $\tau_M$.
%(persistence time).
%In other words, the angle of the
The angle of the 
$i$th agent's motion is $\xi_i(t_k) \in [0,2\pi]$, and
it changes randomly at discrete time steps $t_{k}$ 
($t_{k+1}-t_k=\tau_M$).
The evolution of the $i$th agent's position is therefore
\begin{equation}
\begin{array}{lll}
 x_i(t_k+\Delta t) &=& x_i(t_k) + v \cos \xi_i(t_k) \Delta t \mod L \\
y_i(t_k+\Delta t) &=& y_i(t_k) + v \sin \xi_i(t_k) \Delta t \mod  L,
\end{array}
\label{eq:position}
\end{equation}
where $\Delta t \le \tau_M$.
The motion of the agents is diffusive, with a diffusion coefficient of
$D\sim v^2 \tau_M$.

%For this reason, we frame our problem in the analogous language of
%oscillator synchronization.
%In this paper the dynamics of the oscillators are based on the Kuramoto model
%\cite{Kuramoto84,abprs05}.
In this paper the dynamics of the oscillators are based on the Kuramoto model \cite{Kuramoto84},
which has been applied to technological 
problems recently %\cite{fnp07,lkph06,klmj08}. 
\cite{fnp07}.
The time evolution of the phase of oscillator $i$ is represented as
%\begin{eqnarray}
%\begin{split}
%\varphi_i (t+\tau_P) &= \varphi_i (t) + \sum_{j=1}^N \sigma(d_{ij}) \sin\left(\varphi_j(t) - \varphi_i(t)\right)
%\label{eq:theta1} \\
%\sigma(d_{ij}) &=
%\begin{cases}
%\sigma & (d_{ij} <d) \\
%0 & (d_{ij} >d)
%\end{cases}
%\end{split}
%\end{eqnarray}
\begin{equation}
\varphi_i (t+\tau_P) = \varphi_i (t) + \sigma \sum_{j,d_{ij}<d} 
%\textcolor{red}{\sum_{j=1}^N}  
\sin\left(\varphi_j(t) - \varphi_i(t)\right)
\label{eq:theta1} 
\end{equation}
where $d_{ij} = \sqrt{(x_i-x_j)^2 + (y_i - y_j)^2}$.
%\textcolor{red}{and $\sigma(d_{ij})=\sigma$ if $d_{ij}<d$ and otherwise
%$\sigma(d_{ij})=0$.}
Since mobile devices emit
signals at discrete time intervals, the individual phases are updated at 
discrete time steps of duration $\tau_P$. 
Only devices within a distance $d$ of each other can interact to approach their phases.
%The physical meanings and values of all parameters are summarized in the
%Supplemental Information.
%{In this paper we  fix $N=100$, $L=200$, $v=10$, and $\tau_M=1$.}

%\textcolor{red}{
When the phase difference is small, 
%the sine function in %(\ref{eq:theta1})
Eq.~(\ref{eq:theta1}) can be well approximated by
% its argument and this approximation
the linearized equation
%In such a case we can write 
\begin{equation}
\varphi_i(t+\tau_P) = \varphi_i(t) - \sigma \sum_{j=1}^N L_{ij}(t) \varphi_j (t),
\label{eq:lin}
\end{equation}
which provides interesting hints about the dynamical behavior.
%which is the discrete version of the diffusion equation.
$L_{ij} (t)= [k_i(t) \delta_{ij}-c_{ij}(t)]$ is the
{time dependent} Laplacian matrix {with}
%we have defined %the time dependent connectivity matrix
$c_{ij}(t)=1$ if $d_{ij}<d$ ($i\neq  j$) and $c_{ij}(t)=0$ otherwise.
$k_{i}(t)$ represents %the degree of the $i$th node, i.e.
the number of oscillators that are around $i$ within a range $d$.
%Then, we can conclude that $T$ is also a good estimate for the
%time to synchronize.
%}

We first note that the
instantaneous coordination of agents %in a time-varying topology
has statistical properties similar to  those of a 
continuum percolation.
Simply changing $d$ can be 
enough for the system to enter a
different topological (static) configuration.
 The transition takes place at $(N-1) \pi d_c^2 /L^2 \approx 4.51$ 
(see \cite{dc02} and references therein).
%(see \cite{dc02,bbw05} and references therein);
%for our choice of parameters, $d_c \approx 24.1$. 
In this paper we  fix $N=100$, $L=200$, $v=10$, and $\tau_M=1$, which
imply $d_c \approx 24.1$.
%In our scenario,
Since agents move and the system is 
%the maximum distance for communication is
finite, we do not 
observe a sudden 
transition %at the particular values predicted by the static theory. 
at $d=d_c$.
%; rather,}
%we expect \textcolor{red}{a global} dynamical behavior of the system to change
%gradually around %
%{$d_c$.}
%
%\textcolor{blue}{$\leftarrow$ two paragraphs combined}
%For very small $d$, the network consists of isolated units. For intermediate values 
%of $d$, the system forms clusters (groups of
%connected agents). Generally speaking, as $d$ increases the size of the clusters also
%increases, until a single cluster spanning the whole system appears.

%In an infinite system, this point is called the {\em continuum percolation
%transition}.
%For much larger values ($d\approx L$),
%a complete graph appears.

%Starting from random initial phases and positions, we let the system evolve according to
%Eqs.~(\ref{eq:position}) and (\ref{eq:theta1}). Figure~\ref{fig:movies}
%presents time sequences of four snapshots for three different parameters.
%Synchronization
%emerges through motion and intermittent communication between agents,
%even though a single connected component never forms  below the continuum percolation threshold
%(Fig.~\ref{fig:movies}(a) and (b)).
%Above the percolation threshold (Fig. \ref{fig:movies}(c)), agent motion is not necessary but still
%helps the system reach its final state.

\begin{figure}[ht]
\begin{minipage}{0.48\textwidth}
\includegraphics[width=0.24\textwidth]{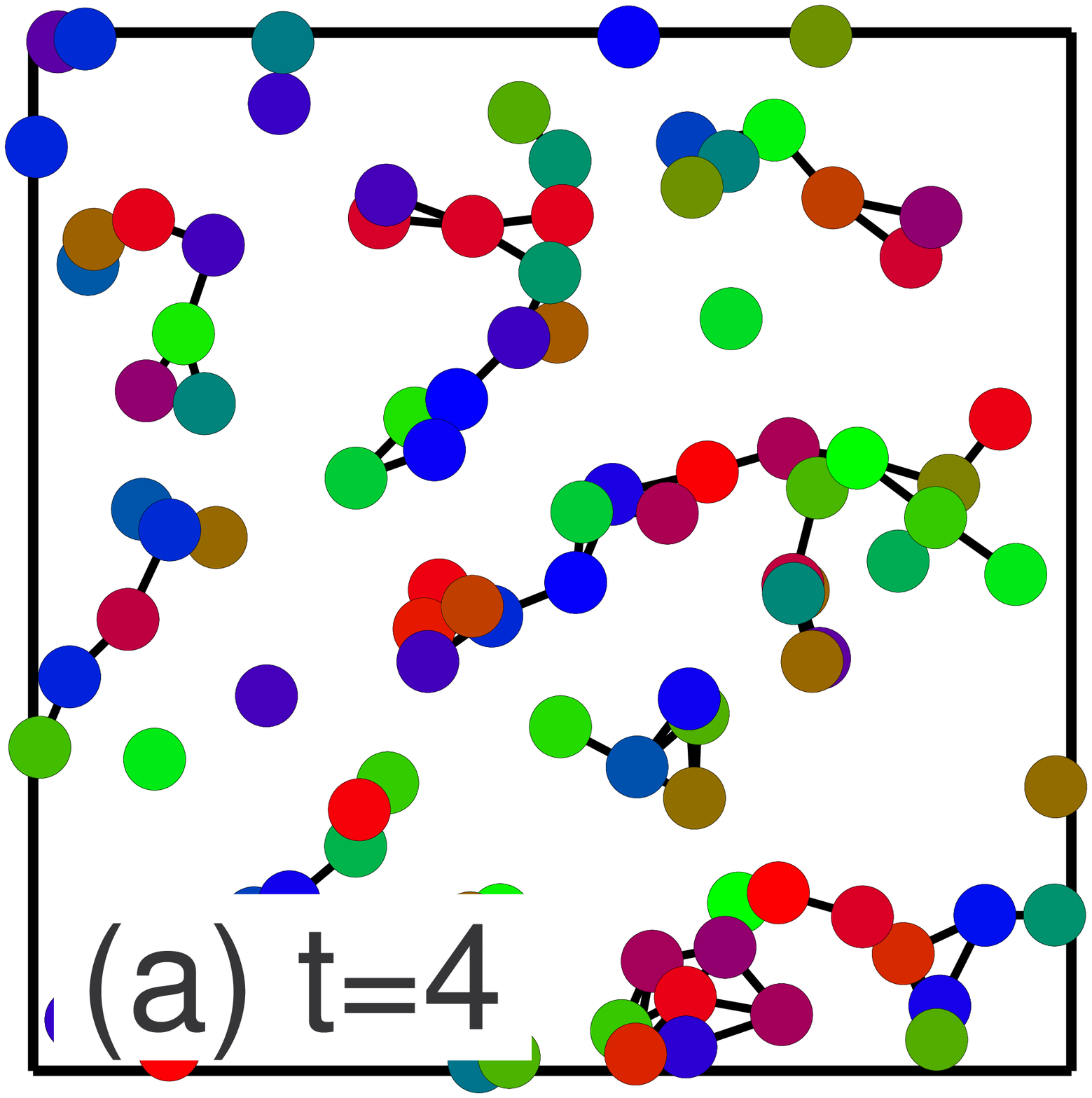}
\includegraphics[width=0.24\textwidth]{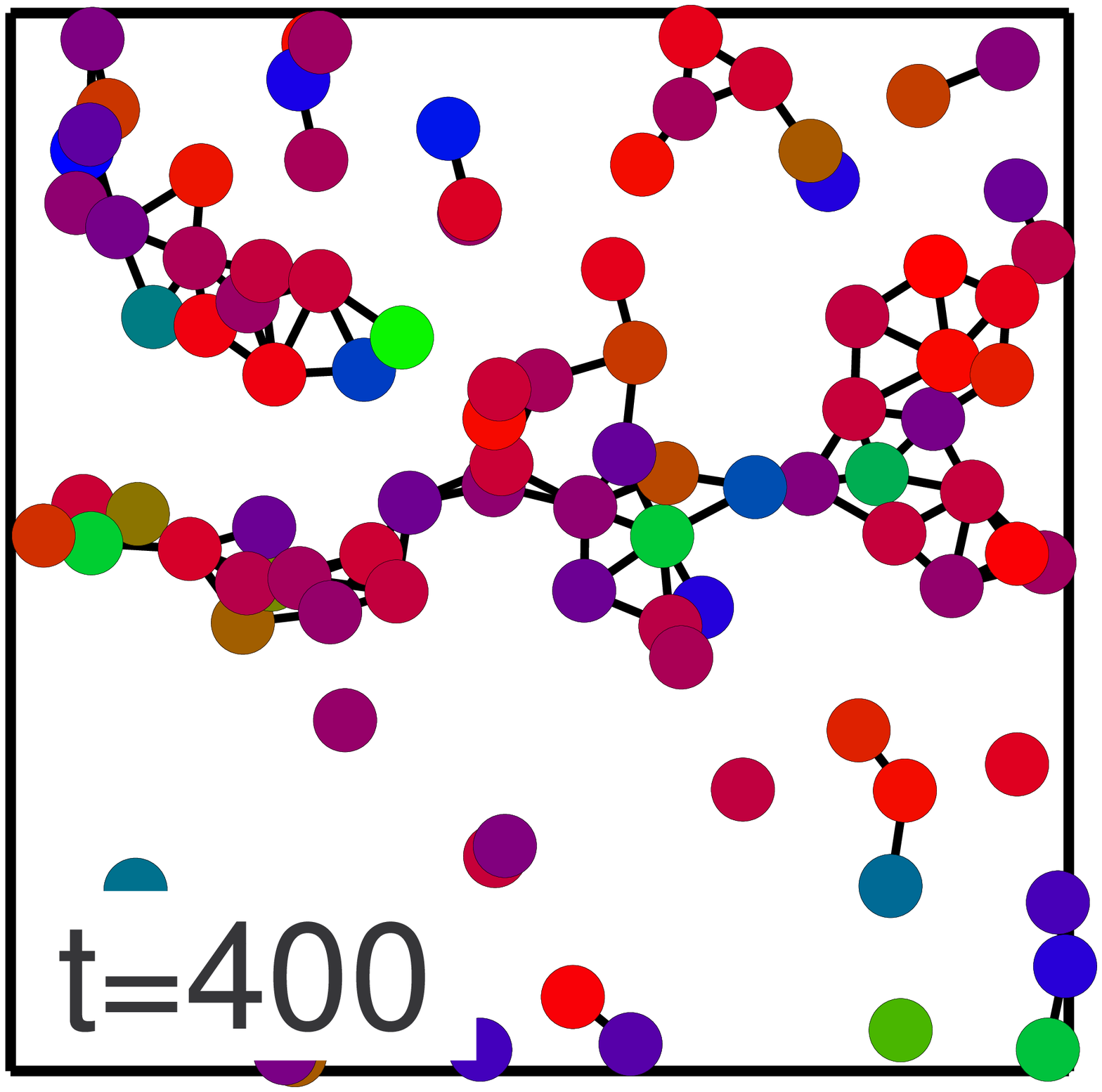}
\includegraphics[width=0.24\textwidth]{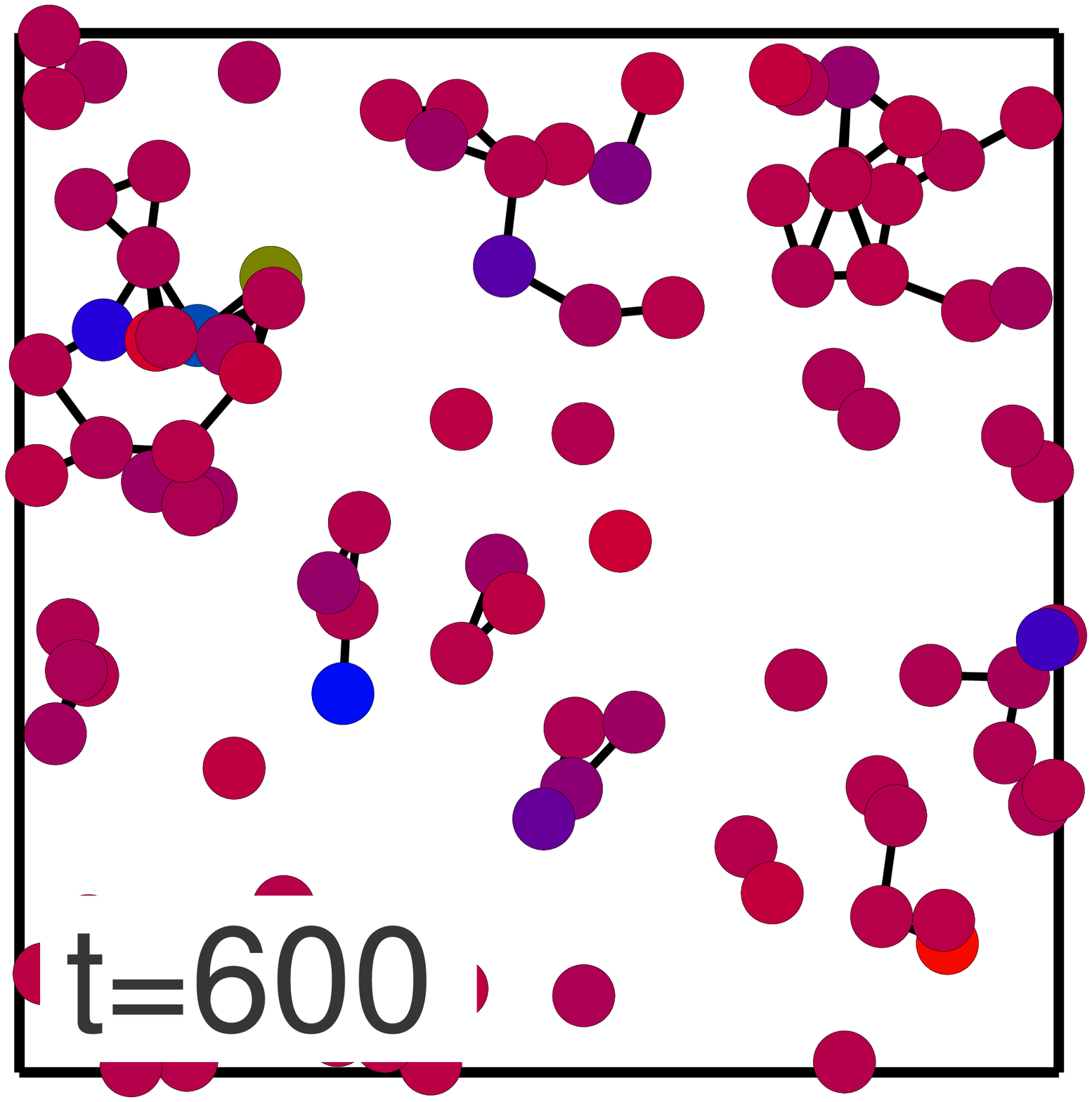}
\includegraphics[width=0.24\textwidth]{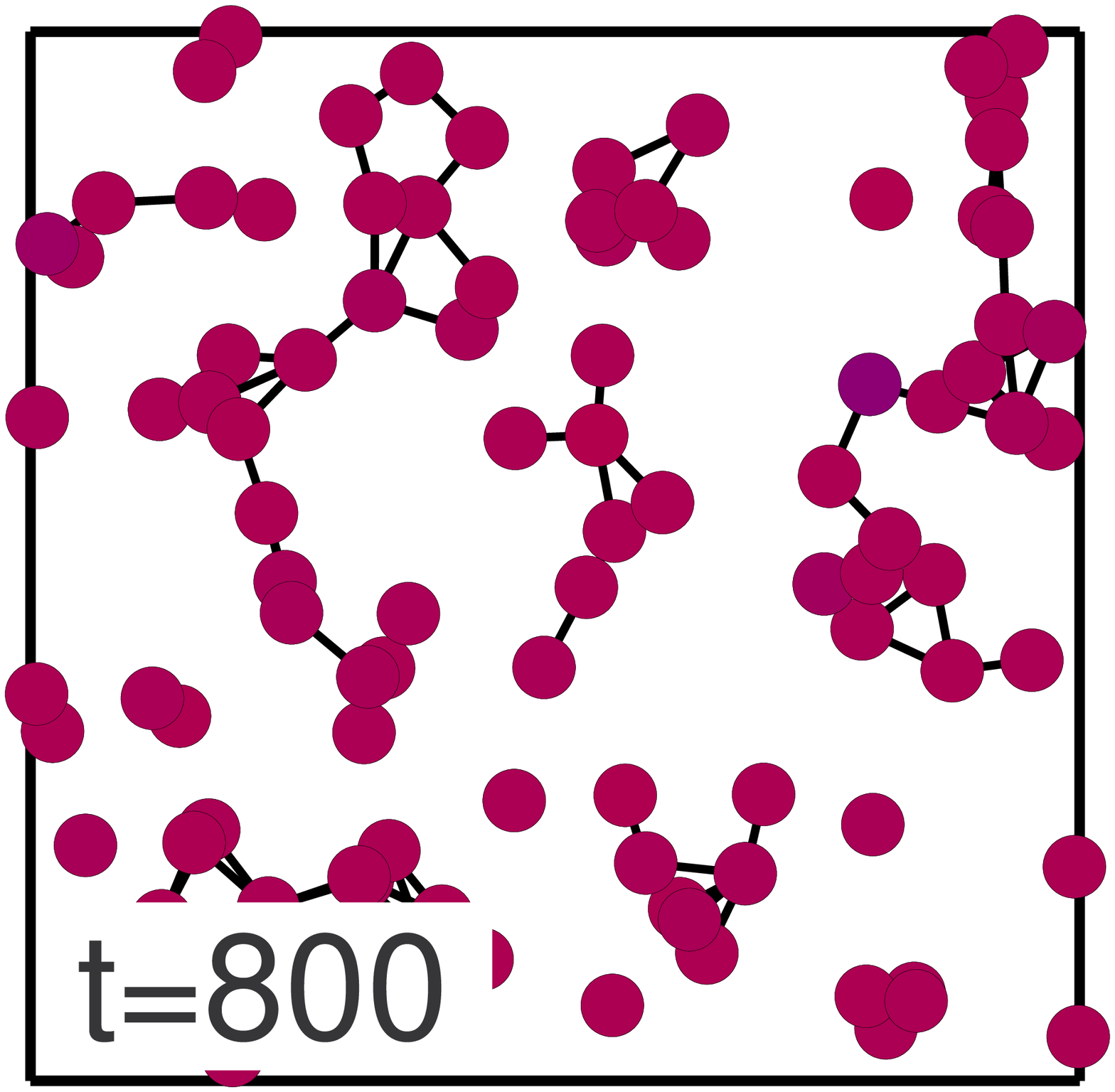}
\\
\includegraphics[width=0.24\textwidth]{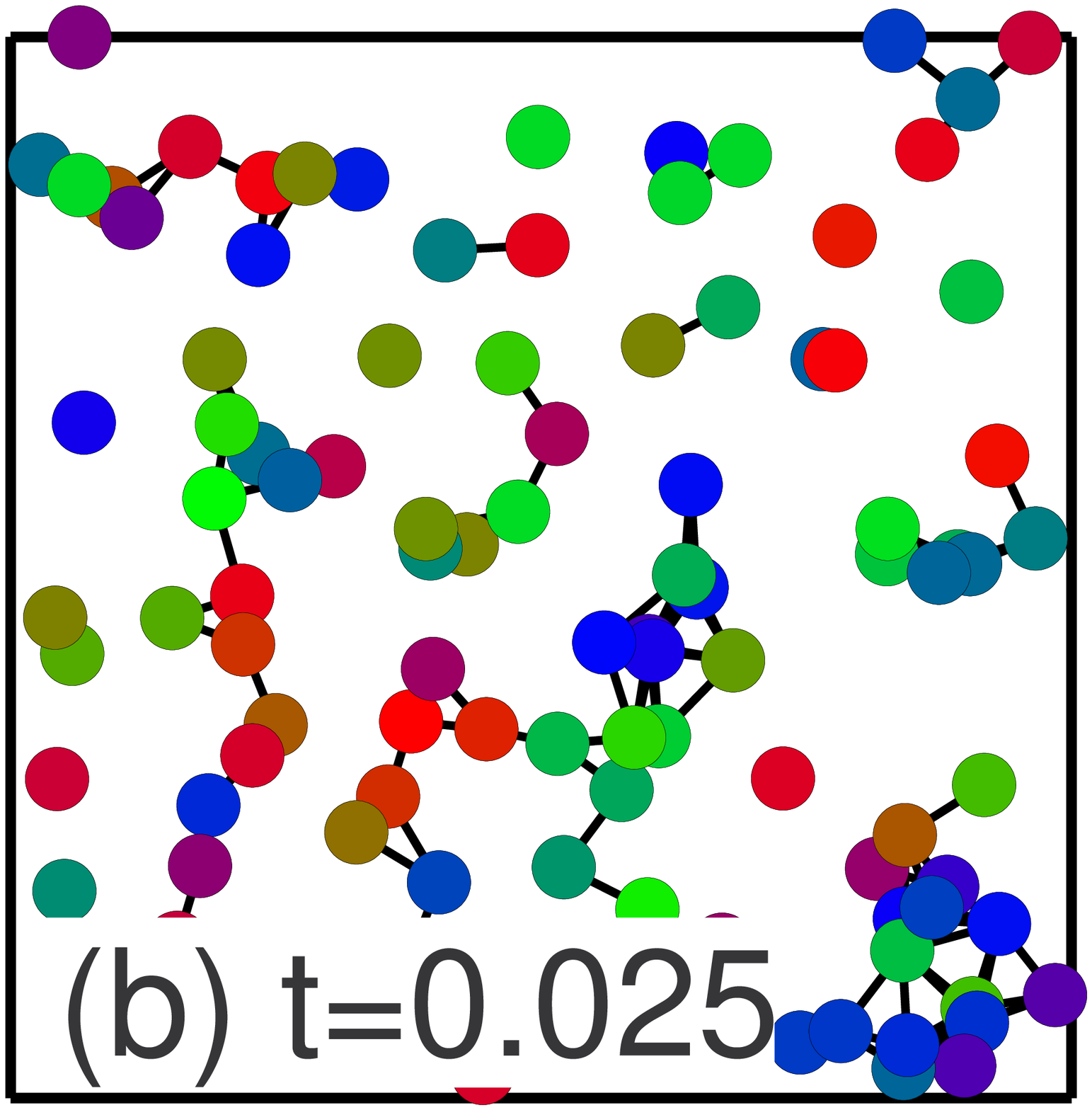}
\includegraphics[width=0.24\textwidth]{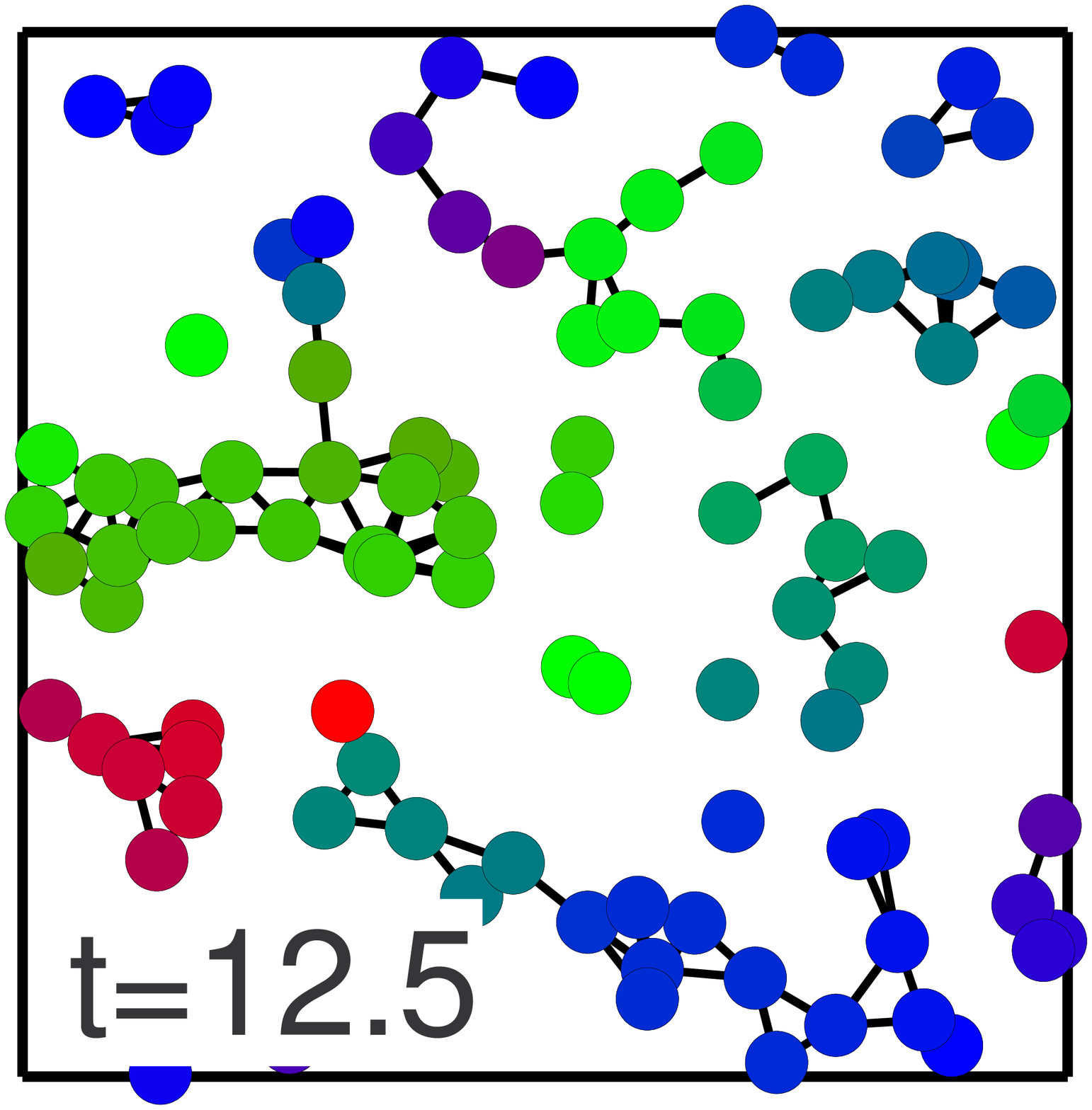}
\includegraphics[width=0.24\textwidth]{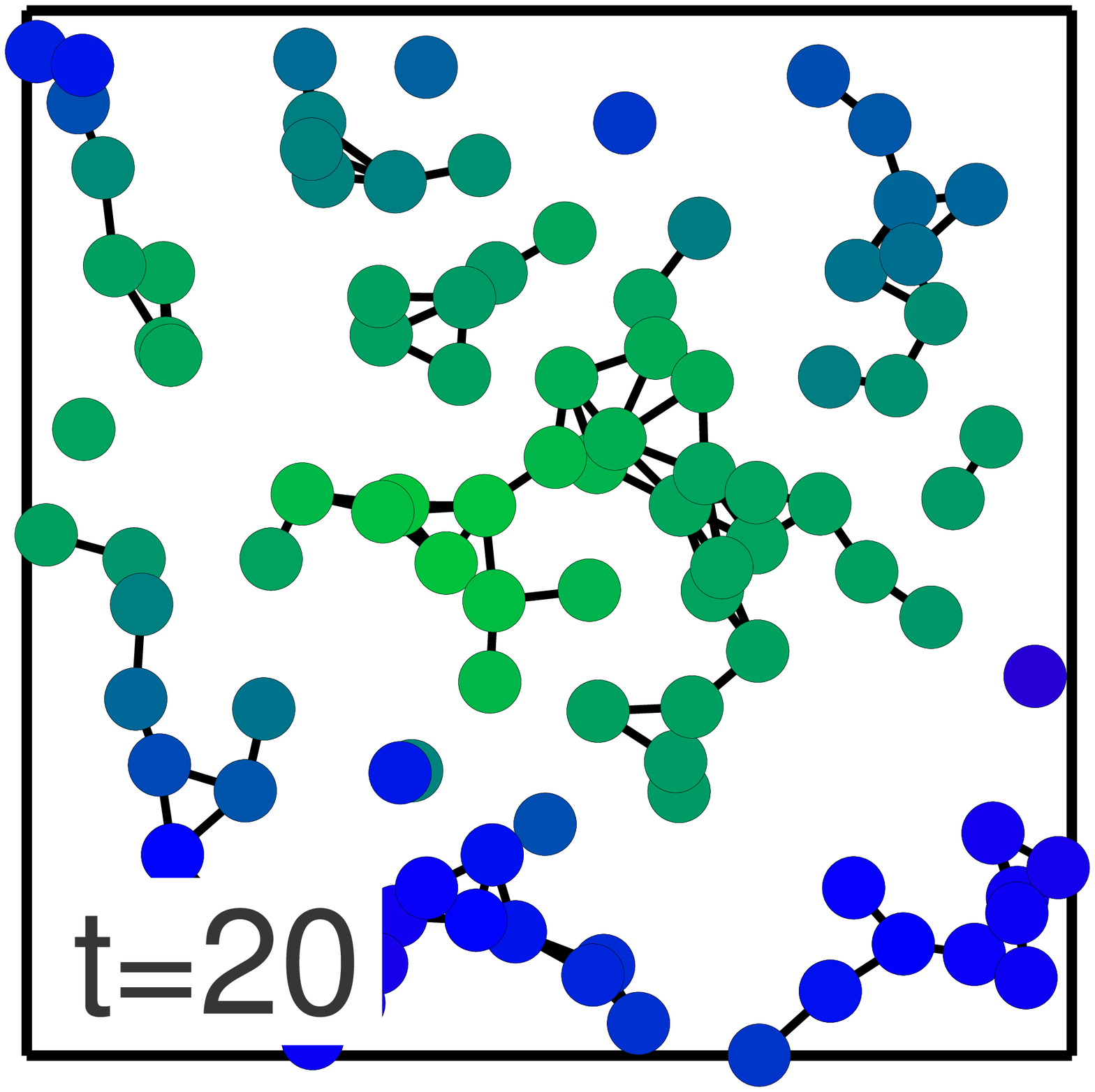}
\includegraphics[width=0.24\textwidth]{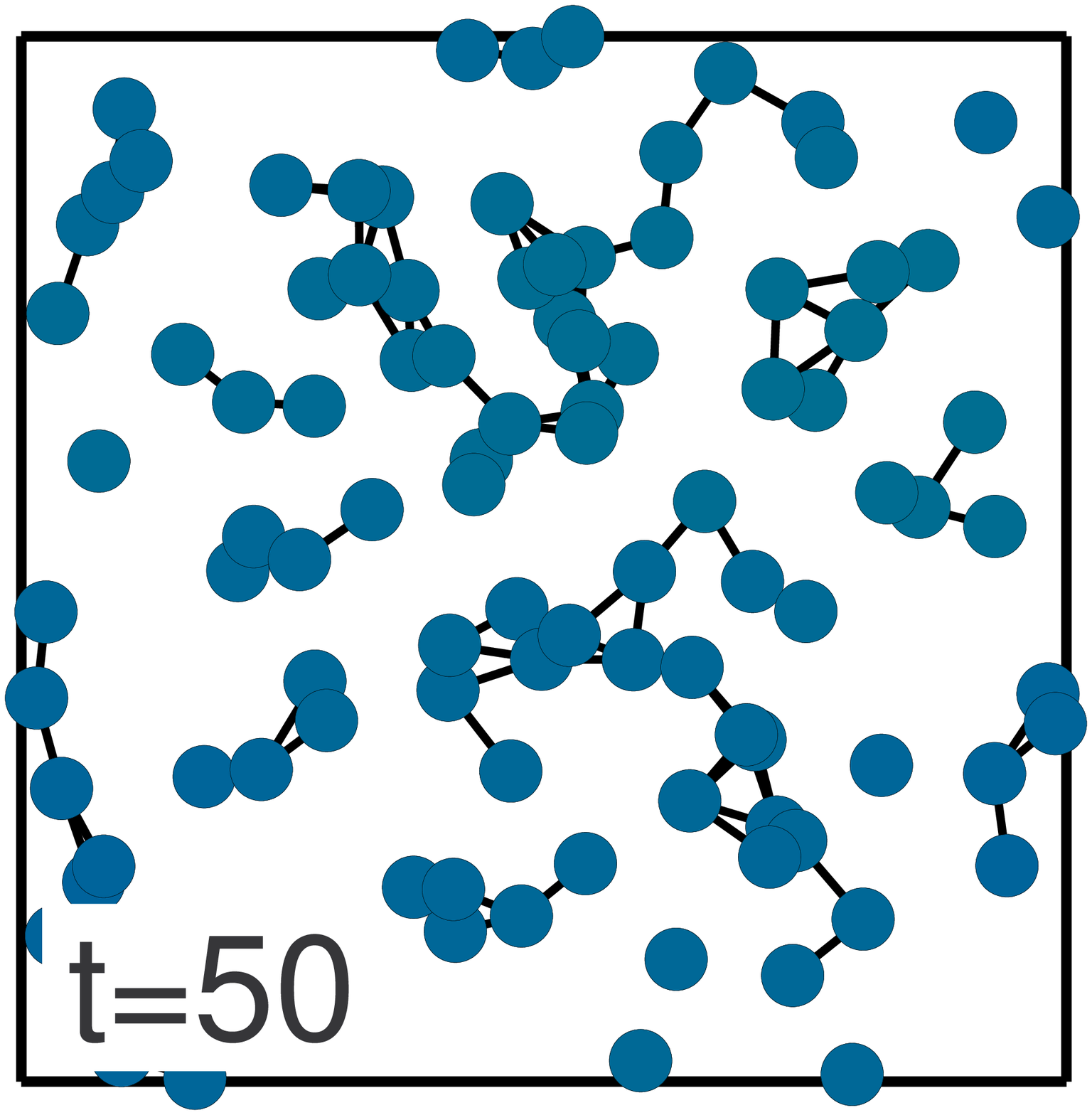}
\\
\includegraphics[width=0.24\textwidth]{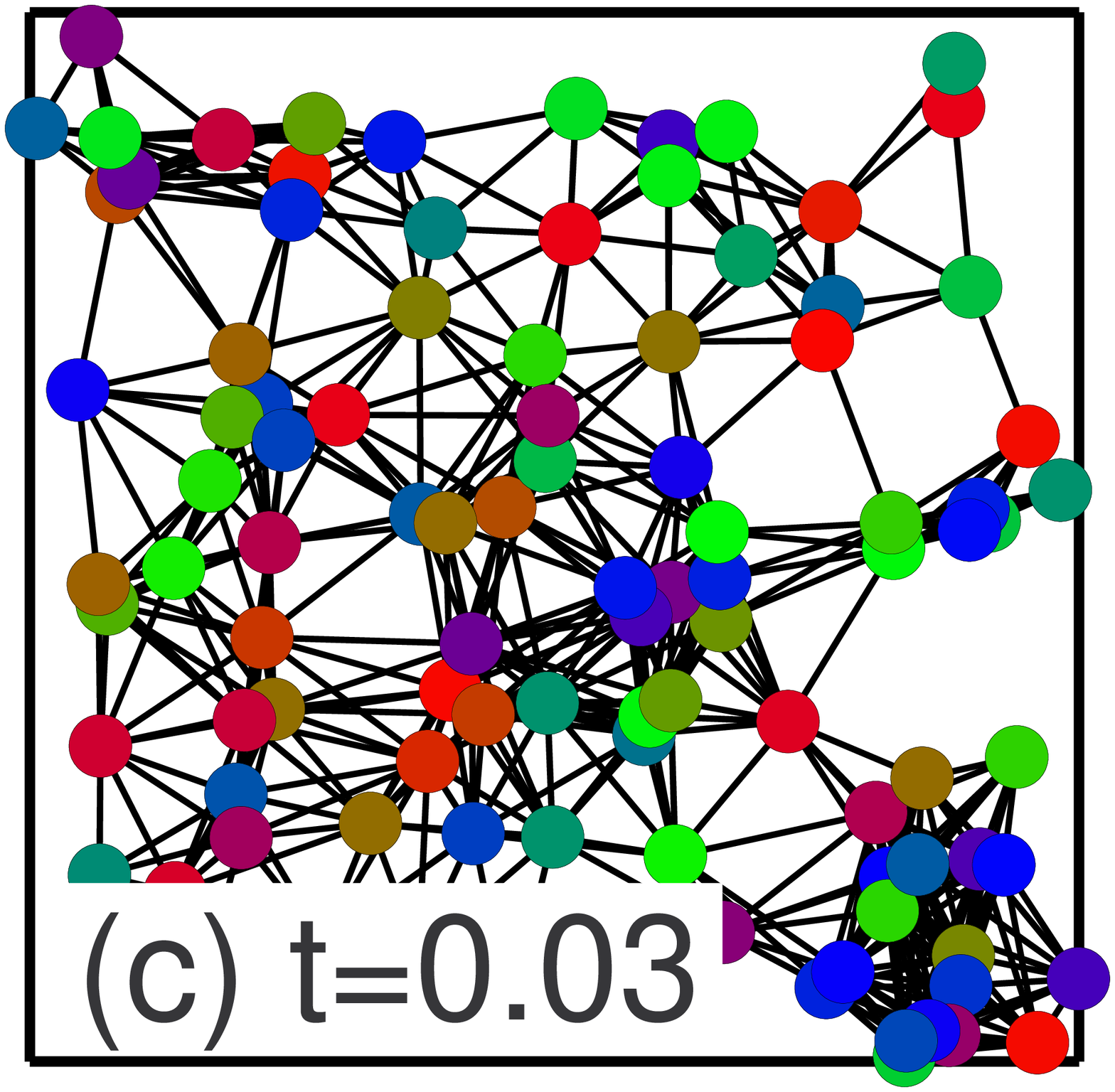}
\includegraphics[width=0.24\textwidth]{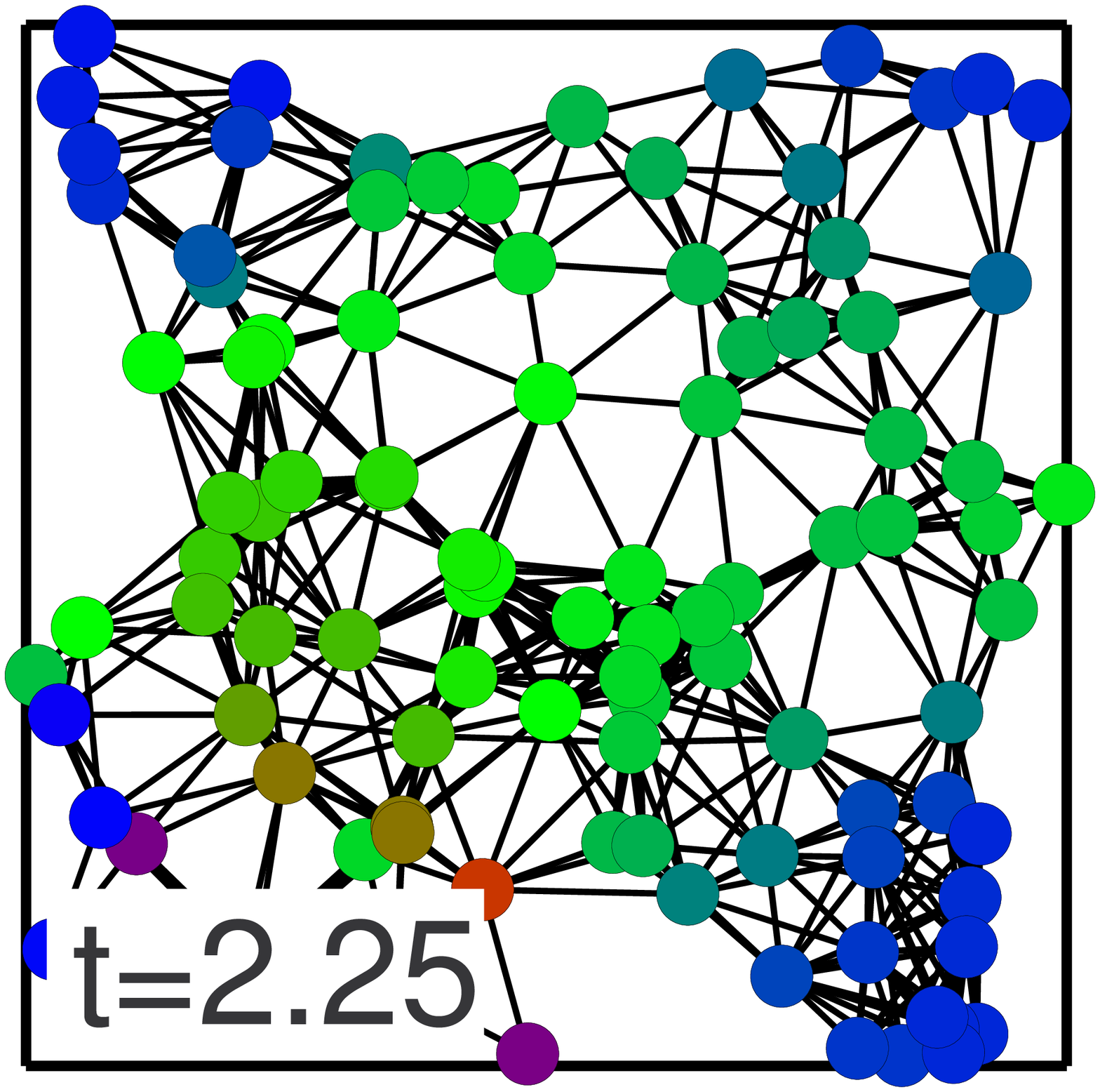}
\includegraphics[width=0.24\textwidth]{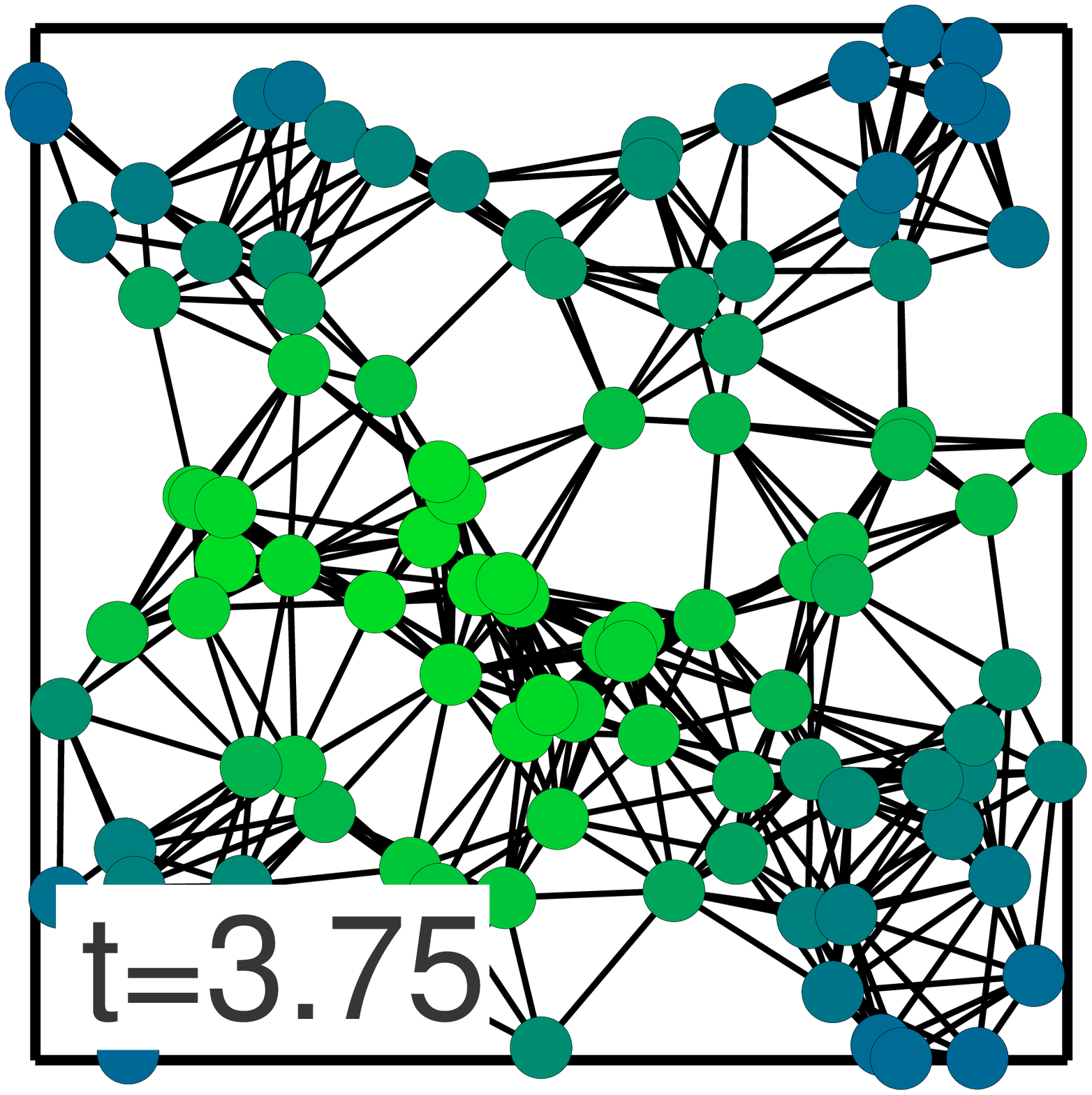}
\includegraphics[width=0.24\textwidth]{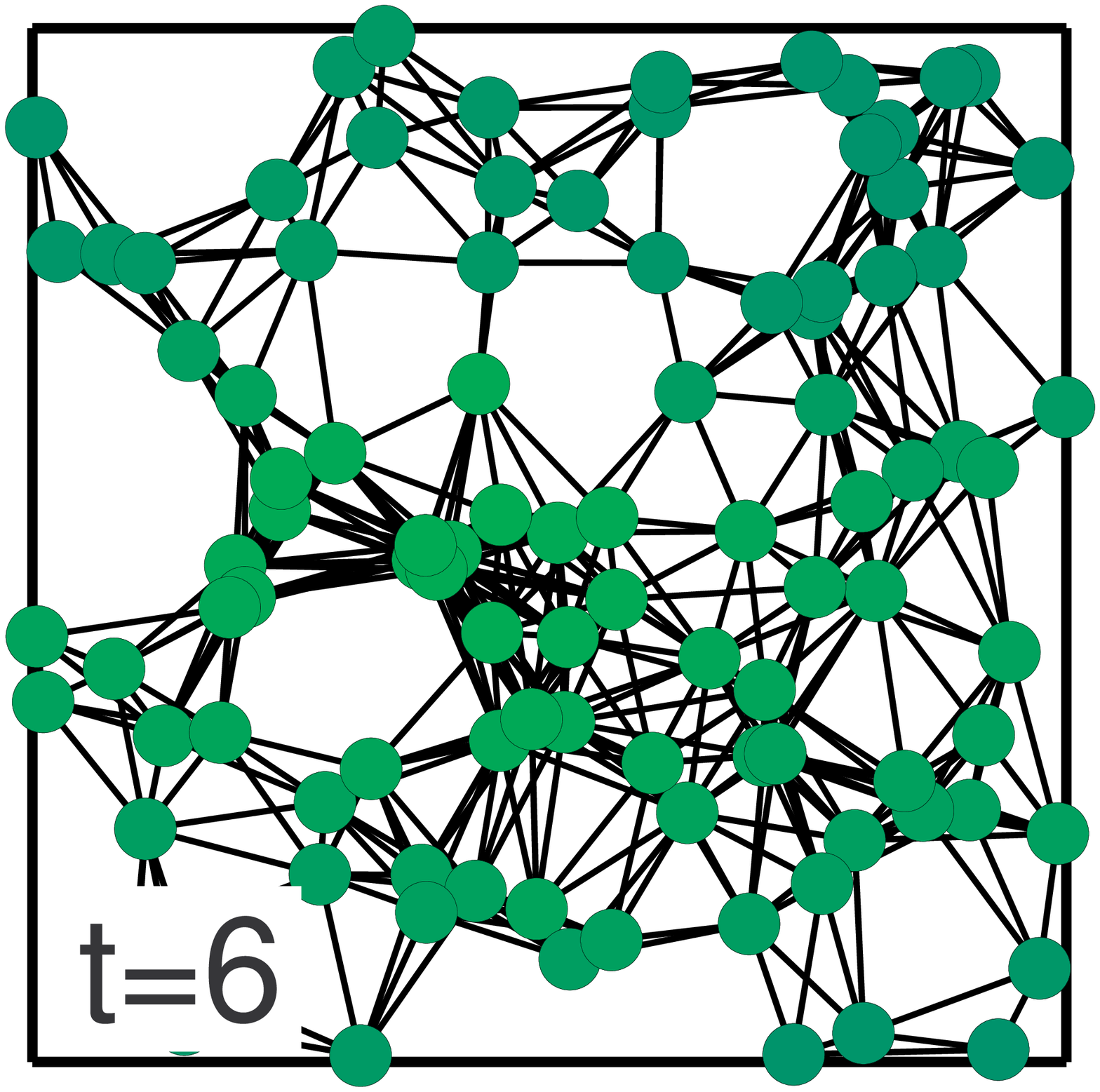}
\caption{ \label{fig:movies} 
(Color online) Time evolution of 
%%the agents
%% Each graph displays 
the positions and phases of the agents.
%% in one of our simulations at a single point in time.
The colors are mapped to phases on the interval $[0,2\pi]$,
lines are drawn between two nodes when their separation is less than $d$.
Each row corresponds to a different set of 
parameters:
%, and viewed from left to right
%%%depicts the evolution of a simulation. 
%All three simulations use the parameter $\sigma=0.005$.
%%%In the top and middle rows, the interaction range $d$ is small. Both systems are therefore 
%%%below the percolation threshold, and give rise to
%%%multiple clusters.
%Other parameter values are
$\tau_P$=1, $d=20<d_c$ for (a), 
$\tau_P$=0.01, $d=20$ for (b), and  $\tau_P$=0.01, $d=40>d_c$ for (c).
%%Early in the simulation, the oscillators are desynchronized ({\bf a}, $t=4$). 
%%Even in the second stage ({\bf b}, $ t=400$) there is no evidence of local 
%%synchronization. 
%%At late times, all oscillators are synchronized even though a
%%single connected component never formed ({\bf d}, $t=800$).
%%At late times, all oscillators are synchronized 
%%({\bf h}, $t=50$).
%%Bottom row:
%%This system is above the percolation threshold, so there is a single connected cluster.
%%The system can therefore become synchronized even if the agents do not move.
%%The oscillators are initially desynchronized ({\bf i}, $t=0.03$). 
%%Local synchronization appears early ({\bf j}, $t=2.25$), and the synchronized 
%%regions expand quickly ({\bf k}, $t=3.75$). 
%%Finally, all oscillators are synchronized ({\bf l}, $t=6$).
}
\end{minipage}
\end{figure}

Starting from random initial phases and positions, we let the system evolve according to
Eqs.~(\ref{eq:position}) and (\ref{eq:theta1}) \cite{anime}.
Figure~\ref{fig:movies}
presents time sequences of four snapshots for three different sets of
parameters.
Synchronization
emerges through motion and intermittent communication between agents,
even though a single connected component never forms  below $d_c$
(Figs.~\ref{fig:movies}(a) and (b)).
Above $d_c$ (Fig.~\ref{fig:movies}(c)), agent motion is not necessary but still
helps the system to reach its final state.

%In order to quantify the dynamics towards the synchronized state, we
%\textcolor{red}{In this paper,} we measure the average phase difference
%\begin{equation}
%\end{equation}
%
%which is naturally related to the decay of the normal
%modes, as shown by previous studies of synchronization in complex networks of
%identical oscillators \cite{adp06a,adkmz08}. 
%
%
%The fact that the average phase difference decays exponentially
%The fact of the exponential decay of $\langle \Delta \varphi \rangle$}
%allows us to define the 
In all our simulations, %we find that 
the average phase difference
$\langle \Delta \varphi  \rangle \equiv 
\sqrt{\frac{2}{N(N-1)} \sum_{j<k} 
(\varphi_j - \varphi_k)^2 
}$
decays exponentially after an initial  transient.
We can define  then a {\it characteristic time} $T$ 
in such a way that  $\langle \Delta \varphi \rangle \propto e^{-t/T}$%. 
and  estimate $T$ 
by fitting the numerical data of $\langle \Delta \varphi\rangle$.
Additionally,
$n_T \equiv T/\tau_P$ stands for the number of phase updates the system needs 
to reach complete synchronization, and its inverse defines the system efficiency.
Minimizing $n_T$ leads to more 
efficient use of the mobile devices' batteries.

In the literature related to interacting units in time-dependent 
networks, authors often use 
the  fast-switching  approximation (FSA) 
\cite{bbh04b,frasca08,psbs06,sbr06}.
FSA assumes that the topology changes fast enough, and 
 the entries in the connectivity matrix are replaced by the probability that two units
are within the interaction range under completely random motion 
($\rho=\pi d^2 /L^2 $, for $ d\le L/2$).
The characteristic time $T_{FS}$ within FSA is expressed as 
\begin{equation}
T_{FS} =- \tau_P/ \log [1-\sigma (N-1) \rho].
\label{eq:fs_eff}
\end{equation}
%where
%$\rho$ is the probability that two agents are connected under completely
%random motion
%.
%\textcolor{red}{(see the Supplemental Information for the derivation).}
% The validity of FSA 
% is one of the main subjects of this study.
%Note that t
The effect of agents motion is averaged out, and 
the parameters $v$ and $\tau_M$ do not appear in Eq.~(\ref{eq:fs_eff}).
It is important to note that it makes sense 
only when the time scale of network variations
is much shorter than that of the interaction.

\begin{figure}[h]
%%%\begin{minipage}{0.48\hsize}
%%%\includegraphics[width=\linewidth]{fig2a.eps}
%%%\includegraphics[width=\linewidth]{fig2b.eps}
%\includegraphics[width=0.75\linewidth]{fig2a.eps}
%\includegraphics[width=0.75\linewidth]{fig2b.eps}
%%%\end{minipage}
%%%\begin{minipage}{0.48\hsize}
%%%\includegraphics[width=0.9\linewidth]{fig3.eps}
%%%\end{minipage}
\begin{minipage}{0.49\hsize}
\includegraphics[width=0.98\linewidth]{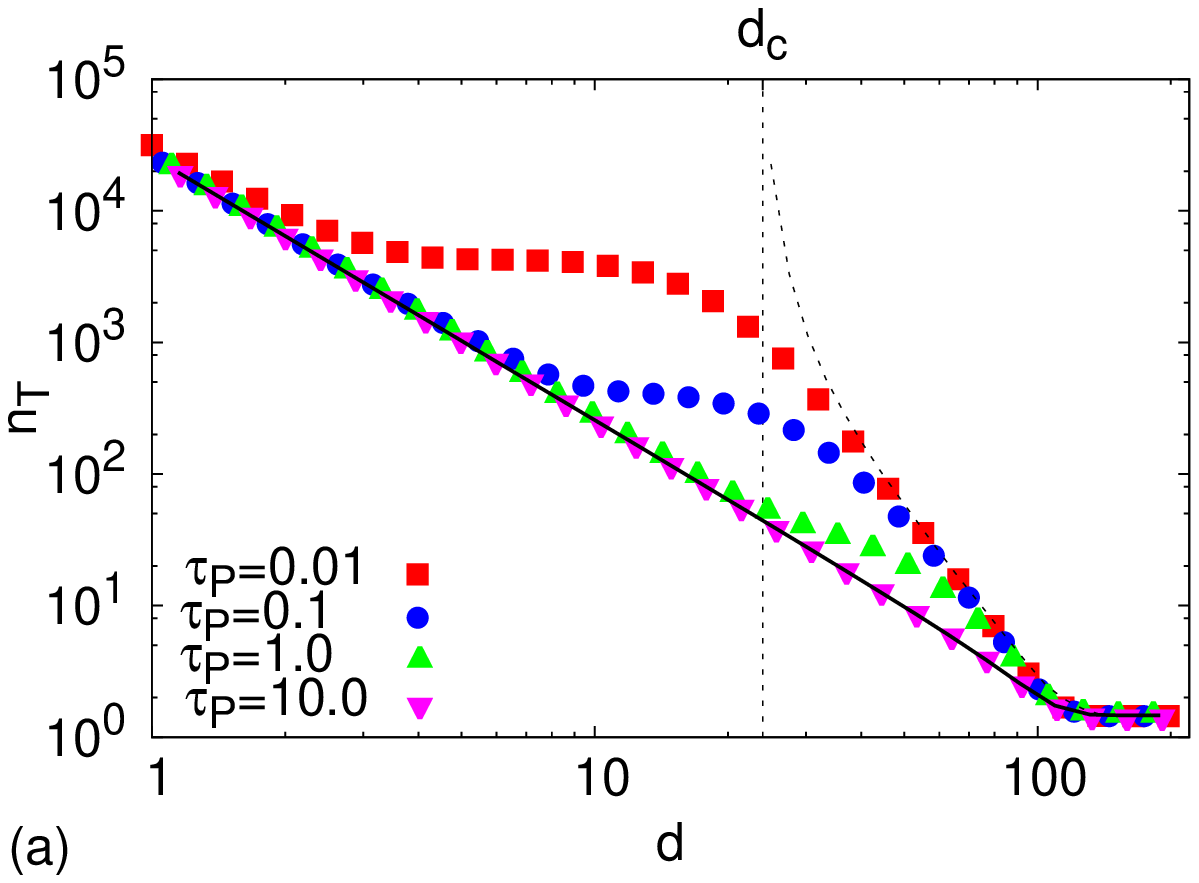}
\includegraphics[width=0.98\linewidth]{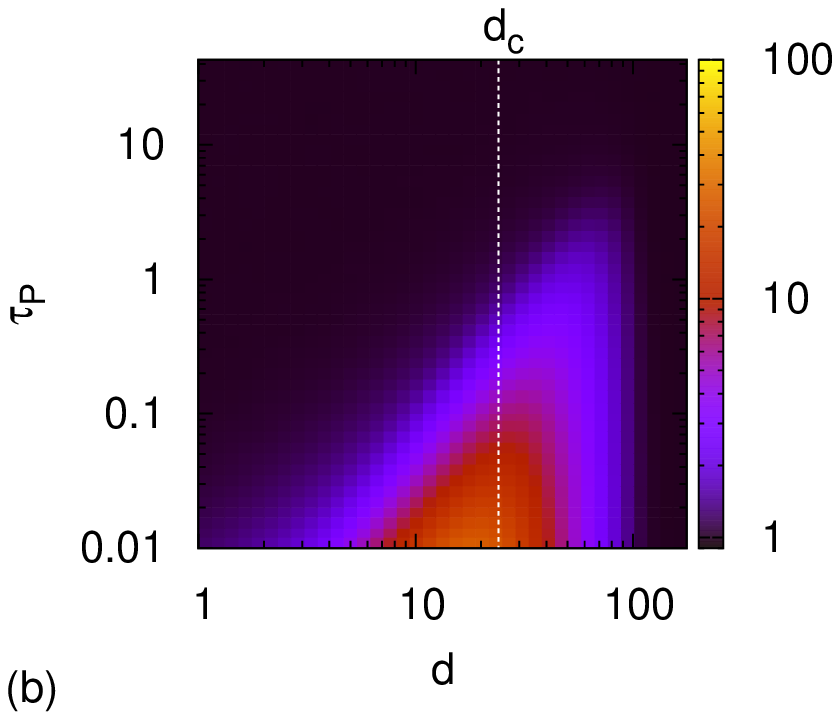}
\end{minipage}
\begin{minipage}{0.49\hsize}
\includegraphics[width=0.98\linewidth]{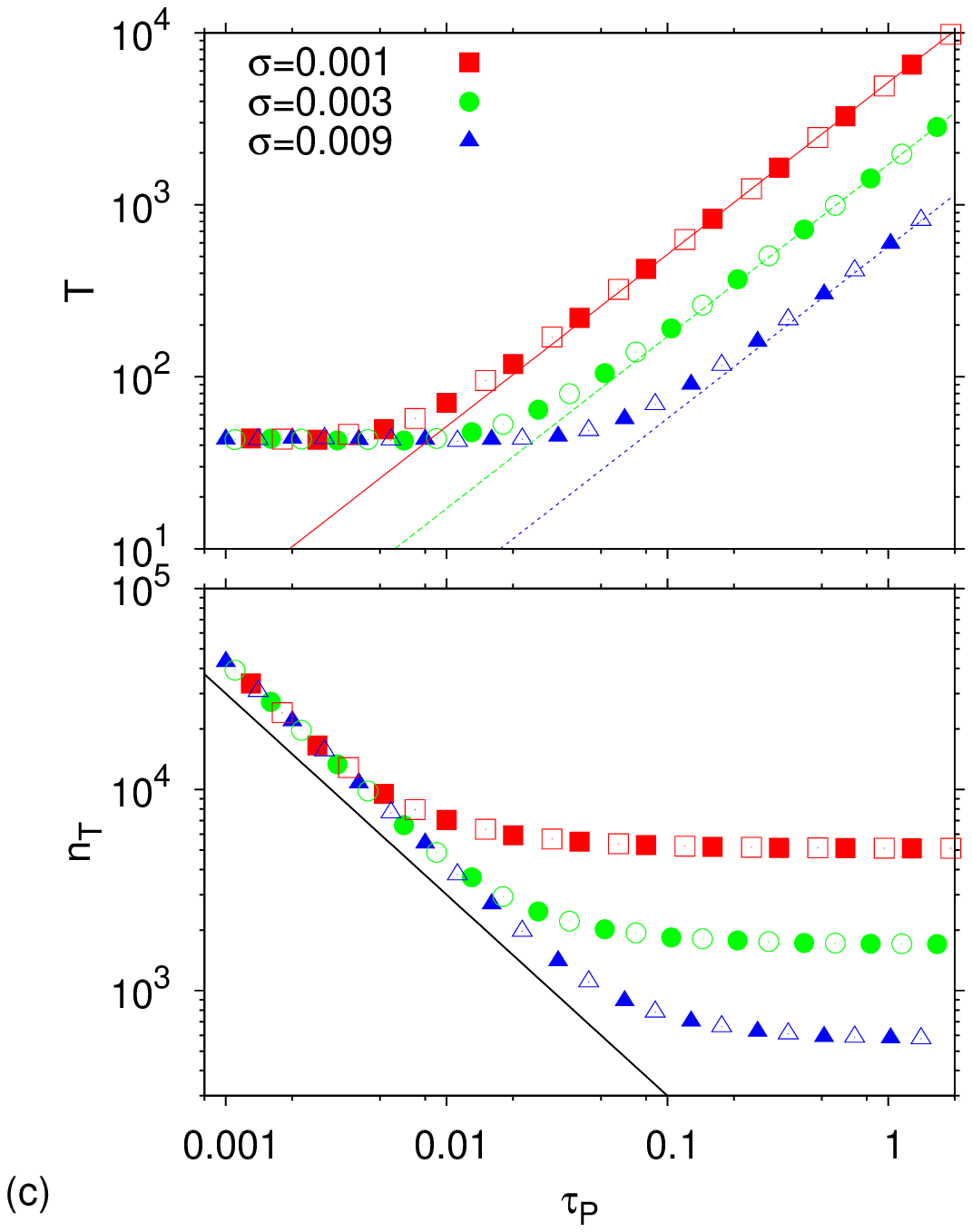}
\end{minipage}
\caption{\label{fig2} 
%\label{fig3}
%{Dependence of the characteristic time on $d$ and $\tau_P$.}
(Color online)
(a) Number of updates required to synchronize, $n_T=T/\tau_P$,
as a function of $d$.
The solid line indicates FSA, $T_{FS}/\tau_P$ (Eq.~(\ref{eq:fs_eff})). The dotted
line is $T_{SL}/\tau_P$ {(Eq.~(\ref{eq:T_sl})).}
%, where $T_{SL}$ is in Eq.~(\ref{eq:T_sl}).
% infinite system \cite{dc02,bbw05}.
(b) The ratio $T/T_{FS}$ in the $\tau_P$-$d$ plane.
%$T$ is obtained by averaging an ensemble of 124 simulations with different initial conditions. 
In (a) and (b) $\sigma=0.005$,
and $d_c$ represents the continuum percolation threshold for an
 infinite system \cite{dc02}.
{(c) 
%Characteristic time $T$ and inverse efficiency $n_T$ as functions of
$T$ and $n_T$ versus $\tau_P$ with different values of $\sigma$
for {$d=5$}.}
Filled and open symbols of the same color
are based on Eq.~(\ref{eq:theta1}) and
the matrix product formulation 
(\ref{eq:theta-evolution}), respectively.
 The colored lines in the top panel 
represent $T_{FS}$ for each case.
The black line in the bottom panel represents $\tau_P^{-1}$.
}
%\textcolor{blue}{mark $d_c$ in (b)}}
\end{figure}

Figure \ref{fig2}(a) plots %the number of signals required for synchronization 
$n_T$ 
as a function of $d$ for various values of $\tau_P$. As intuitively expected, 
there is a decreasing monotonic relationship between $n_T$ and $d$. 
We have also drawn a reference line along $T_{FS}/\tau_P$, which depends
only on $d$. 
For large $\tau_P$ FSA is very accurate over a wide range of $d$, 
since the above mentioned condition for FSA is satisfied.
However, we also identify %observe
 a region in Fig.~\ref{fig2}(a) 
where FSA does not hold 
for intermediate values of $d$. 
The size of this region increases as $\tau_P$ decreases. 

It is important to note that these deviations from FSA take place close to
the continuum percolation transition point.
This is not a sharp transition; instead, $T$ gradually deviates from FSA.
To gain a broader understanding, 
we plot $T/T_{FS}$ over the $\tau_P$-$d$ plane
in Fig.~\ref{fig2}(b). %Along the top of this figure, where
For large enough $\tau_P$, FSA
is very good irrespective of $d$. 
%In this region the quality of the approximation does not depend on $d$.

In order to qualitatively explain the deviations from FSA, 
we consider two characteristic time scales: 
%the interval between phase interactions and the motion of the agents.  
%Broadly speaking, they can be understood as 
%the time scales 
{one}
for clusters to synchronize and %the 
another one for  breaking apart.
The number of time steps (measured in units of $\tau_P$) 
for a cluster to synchronize is, to first order in $\sigma$,
%\be
%n_s=\frac{1}{\sigma \lambda_2^c(d)}.
%\ee
$ n_s=1 / \sigma \lambda_2^c(d)$.
Here $\lambda_2^c$ stands for the smallest nonzero eigenvalue 
of the Laplacian of the cluster.  %%%,
On the other hand, the number of steps %needed 
for a single oscillator to 
leave a cluster of size $\xi(d)$ (an increasing
function of $d$ if $d<d_c$) is
%\be
%n_m=\frac{\xi^2(d)}{v^2\tau_M\tau_P}\ .
%\ee
$n_m=\xi^2(d) / v^2\tau_M\tau_P$.
We can then introduce the ratio
\begin{equation}
\eta= \frac{n_m}{n_s}=\frac{\sigma f(d)}{v^2\tau_M\tau_P},
\label{eq:eta}
\end{equation}
which gives us the dominant time scale,
%which time scale is dominant,
where $f(d)\equiv \xi^2(d) \lambda_2^c (d)$. 
%For a qualitative analysis, it suffices to realize 
Note that 
the topological parameter $d$
(also $N$ and $L$, if they were changed)
appears only in $f$, while the parameters related to agent dynamics appear
only in the denominator.

It is clear from Eq.~(\ref{eq:eta}) that $\eta$ decreases if we increase
$\tau_P$. 
%Therefore, our qualitative relations 
This fact predicts a transition 
in the dominant time scale as we change 
$\tau_P$ which is confirmed in Fig.~\ref{fig2}.
 We expect the same transition in $f(d)$
by changing $d$. Our numerical results in Fig.~\ref{fig2} 
suggest that $f(d)$ is an
increasing function of $d$ well below $d_c$.
In the following, we analyze three different asymptotic behaviors in detail:

i) The condition $\eta {\ll} 1$ 
holds for small $d$ and large $\tau_P$ 
 in Fig.~\ref{fig2}.
In this region, the displacement of agents between $\tau_P$ is large.
Thus, the network connectivity changes very fast 
%and oscillators cannot reliably 
before agents synchronize with their neighbors. 
Figure \ref{fig:movies}(a) shows the evolution of a system under these conditions.
All nodes in the system approach complete synchronization
at approximately the same rate, and non-synchronized nodes may become
spatially isolated.
 We call this mechanism 
{\em global synchronization}.
FSA is accurate for systems of this type.
%(large values of $\tau_P$ and small $d$ in Fig.~\ref{fig2}(b)).
%
%\textcolor{red}{ The parameter
%$v^2 \tau_M \tau_P$
%is large, so consecutive updates are uncorrelated.
%Rather, synchronization is global and appears gradually with some local
% inhomogeneities ({\bf c}, $t=600$). 
%}

ii) Starting from the previous case, $\eta$ increases when either $d$ is increased 
or $\tau_P$ is decreased. 
For $d\ll d_c$,
% is well below the percolation threshold $d_c$, 
$\eta > 1$ implies
that the number of time steps required for agent motions to rewire disconnected
clusters is larger than that %the number of steps 
required for
synchronization inside an isolated cluster. 
Local clusters therefore synchronize very easily before the topology changes
%, as depicted in
{(Fig.~\ref{fig:movies}(b)).}
%In the intermediate
%stage ($t=12.5$ and $t=20$), clusters are internally synchronized.}
The dynamics towards complete synchronization is limited by the 
motion of the agents,
and we call this mechanism {\em multiple cluster local synchronization}.
% The clusters split and merge,
%so complete synchronization will eventually develop. 
%Another feature of this scenario is that
Consecutive positions of agents 
are correlated, and the evolution of the system depends heavily
on the details of the connectivity pattern.
Since FSA neglects such a correlation, it does not properly describe
the synchronization dynamics in this case.
Indeed, as shown in Fig.~\ref{fig2}(b), FSA fails by orders of magnitude.
It is clear from Eq.~(\ref{eq:eta}) that this region is broader if
$\tau_P$ is smaller.
%\textcolor{red}{
% The parameter $v^2\tau_M \tau_P$
% is small, so updates are correlated. 
%Again, the oscillators begin in a desynchronized state ({\bf e}, $t=0.025$). 
%Local synchronization appears very early in the simulation ({\bf f}, $t=12.5$). 
%}

iii) For $d\gg d_c$
%When $d$ is well above $d_c$
 and $\eta > 1$ (implying small $\tau_P$), the whole network is connected ({\em single cluster}). 
In this case, agent motion is not necessary for the final
synchronization.
For a static connected network the characteristic time is $T_{static}=-\tau_P/\log (1-\sigma\lambda_2)$, where $\lambda_2$ is
the second-smallest eigenvalue of the Laplacian matrix \cite{adkmz08}. 
When the network topology changes, its time average
\begin{equation}
%T_{SL} = -\frac{\tau_P}{\langle \log(1-\sigma \lambda_2)\rangle}
T_{SL} = - \tau_P / \langle \log(1-\sigma \lambda_2)\rangle
\label{eq:T_sl}
\end{equation}
is a reasonable upper bound for the characteristic time, since it can only be improved by the motion of the agents. 
Figure \ref{fig2}(a) shows that $T_{SL}$ is not only an upper bound, but also
a good approximation for small values of $\tau_P$.
When $d$ is large enough for the system to form a complete graph, FSA
fits again the numerical result.
% as expected, because all non-zero eigenvalues are identical.

In real {mobile networks}, the signal interval
$\tau_P$ is one of the easiest parameters to control.
Thus, it is important to study %the optimal value of 
{the dependence of $T$ on $\tau_P$ with other %while other
parameters fixed.}
Figure \ref{fig2}(c) describes this dependence. Decreasing $\tau_P$
saturates
% the characteristic time
$T$, while increasing $\tau_P$ saturates %the parameter
$n_T$.
Therefore, there exists an optimal value of $\tau_P$ in the intermediate region
that simultaneously achieves rapid synchronization and high efficiency.
This result demonstrates the importance to take into account the deviation
from FSA, since it causes the saturation of $T$.
% is caused by this deviation.

%\begin{figure}[h]
%\begin{center}
%\includegraphics[width=0.9\linewidth]{fig3.eps}
%\caption{\label{fig3}
%{Characteristic time $T$ and inverse efficiency $n_T$ as functions of
% $\tau_P$ for different values of $\sigma$.}
%Filled and open symbols of the same color
%are based on Eq.~(\ref{eq:theta1}) and
%the matrix product formulation 
%(\ref{eq:T-pm}), respectively.
% The colored lines in the top panel 
%represent $T_{FS}$ for each case.
%The black line in the bottom panel represents $\tau_P^{-1}$.
%Here we fix {$d=5$}.
%\textcolor{blue}{Figures can be made wider and shorter in vertical scale?}}
%\end{center}
%\end{figure}

In order to get some analytical insight,
 we introduce the normal modes of the linear dynamics (\ref{eq:lin}).
%they
%are time dependent because the network topology
% (and hence $L_{ij}$)
%In our current case, though, 
Note that $L_{ij}$ changes with time.
Let $\theta_l(t)$ be the normal modes corresponding to an eigenvalue 
$\lambda_l$ at time $t$ which satisfies 
$\varphi_j(t) = \sum_{l=1}^{N} U_{jl}(t)\theta_l(t)$,
where $U_{jl}(t)$ is the orthogonal matrix 
with a unit determinant.
%of the transformation from the
%original coordinates to the normal coordinates, 
%%being its columns 
%%the normalized eigenvectors of the Laplacian matrix,
%%we can write
%at any time step. 
Multiplying the two sides of Eq.~(\ref{eq:lin}) 
%in the Supplementary
%Information by
by the transpose $U^{T}_{li}(t+\tau_P)$ from the left, 
we get
\begin{align}
{\theta_l}(t+\tau_P) &=
%  & \sum_{i,m}U^{T}_{li}(t+\tau_P)
%U_{im}(t) \left[ 1 - \sigma \lambda_m(t)  \right]
%\theta_m(t) \nonumber \\
%&\equiv &
\sum_{m=1}^N O_{lm}(t) [1-\sigma \lambda_m (t)]\theta_m(t) .
\label{eigenmodes_eq}
\end{align}
Here, $O_{lm} (t)\equiv \sum_{i}U^{T}_{li}(t+\tau_P) U_{im}(t) $
 is orthogonal.
%an orthogonal matrix with a
%unit determinant.
Then after an arbitrary number of time steps we get
\begin{align}
\theta_{l_n}(t+n\tau_P) = \prod_{q=0}^{n-1} 
\left[ \sum_{l_q=1}^{N} O_{l_{q+1} l_q} (1-\sigma \lambda_{l_q}) \right]
 \theta_{l_0}(t),
\label{eq:theta-evolution}
\end{align}
where $l_q$ denotes the suffix corresponding to an eigenmode at time $t+q\tau_P$.
The product of these matrices separately
describes the transformation
of the normal modes of instantaneous networks by
$O_{l_{q+1} l_q}$ and the decay of each eigenmode by $(1-\sigma \lambda_{l_q})$.

This product of $n$ matrices 
in Eq.~(\ref{eq:theta-evolution})
can be diagonalized.
Let its eigenvalues be $(1-\sigma\Lambda_i)^n$ with
$0=\Lambda_1 \le \Lambda_2 \le \cdots \le \Lambda_N$,
the limit of $\Lambda_i$ exists for $n\rightarrow \infty$.
The characteristic time can then be written
as $T=-\tau_P/\log (1-\sigma\Lambda_2)$.
% in terms of the 
%second largest eigenvalue.}
%
%,and its eigenvalues converge \textcolor{blue}{(to what????)}
%for large enough $n$.
%The characteristic time $T$ can then be written in terms of the 
%second largest eigenvalue.
%
%  as $1-\sigma\Lambda_2$ as
%\be
%T=-\frac{\tau_P}{\log (1-\sigma\Lambda_2)}\ .
%\label{eq:T-pm}
%\ee 
Figure \ref{fig2}(c) compares $T$
obtained by this method to that 
directly measured from simulations.
%{based on Eq.~(\ref{eq:theta1}).}
%found in the numerical simulation. 
 Their agreement is excellent, 
even for smaller values of $\tau_P$ where FSA does not hold.
Our procedure can be generalized 
%to describe synchronization dynamics 
to any other evolving network \cite{bhcakp06}.
%\cite{bhcakp06,zhl09},
%within the Master Stability Function formalism
%\cite{adkmz08}.

The two non-trivial behaviors obtained in the simulations
can also be distinguished in the matrix product formulation.
For $\eta \ll 1$, 
%i.e., the time scale of the agents motion
% is much shorter
%than that of the local synchronization,
{where FSA holds,}
 the agents move a sufficiently long distance during $\tau_P$,
and $L_{ij}(t)$ and $L_{ij}(t+\tau_P)$ are regarded as independent.
Thus, their eigenvalues, $\lambda_l (t)$ and
$\lambda_l(t+\tau_P)$ are uncorrelated.
Then we can expect $\prod_{q=1}^n (1-\sigma
\lambda_{l_q}) \approx 
%\exp[ n \langle \log(1-\sigma \lambda) \rangle]$ }
e^{n \langle \log(1-\sigma \lambda) \rangle}$ 
% for every possible combination of $l_q$'s  in
% Eq. (\ref{eq:theta-evolution}) in the Supplementary Information,
if we neglect the fluctuation,
where the bracket represents the average over eigenvalues of the
Laplacian matrix.
%At first, we write $\lambda_{k_q} = \langle \lambda \rangle +
Then we get 
$ T = - \tau_P/\langle \log(1-\sigma \lambda )\rangle$.
%}%
%
%\textcolor{red}{
Since the average eigenvalue of the Laplacian matrix is 
the average degree, we get $\langle \lambda \rangle = (N-1)\rho$.
%where $\rho$ is defined by Eq. (\ref{eq:rho}) in the supplementary information.
Expanding the characteristic time in powers of $\sigma$, we have 
\begin{align}
\tau_P / T = \sigma (N-1) \rho + \mathcal O(\sigma^2),
\label{eq:fsa}
\end{align}
which is equal to $\tau_P/T_{FS}$
up to the lowest order in $\sigma$.

For %the case 
$\eta \gg 1$ and $d \ll d_c$
%, there exist 
(multiple clusters),
there is more than one zero eigenmode of the instantaneous Laplacian matrix,
%and non-zero eigenmodes decay fast enough before the network topology changes.
and $\theta_l (t)$ corresponding to a non-zero eigenvalue
$\lambda_l(t) \neq 0$ vanishes before the topology changes, implying that local synchronization  is achieved.
Hence, the dynamics of the system
is governed by the decay of the zero-eigenmodes caused by 
the topological change.
%The characteristic time of the agent motion is larger than that of 
%the synchronization inside the cluster.
%If $\tau_P$ is small enough, such that $\eta\gg 1$ is satisfied,
%$\theta_l (t)$ corresponding to a non-zero eigenvalue $\lambda_l(t) \neq 0$
%vanishes before the topology changes, implying that local synchronization  is achieved.
Even if we increase 
the number of signals by decreasing %the signal interval
$\tau_P$, we cannot get a further decrease of
the synchronization error between disconnected clusters.
Therefore, we expect that $T$ converges to a finite value for
$\tau_P\rightarrow 0 $.
Since $\sigma$ appears only in the non-zero eigenmodes $(1-\sigma \lambda_l)$ 
%and does not appear in our approximation neglecting the non-zero eigenmodes,
which are neglected in our approximation,
the converged value of $T$ does not depend on $\sigma$ either
 (Fig.~\ref{fig2}(c)).

%For $d \gg d_c$, the whole network is almost always connected
%and $\lambda_2 (t)$ is finite.
%For $\eta \gg 1$,  the motion is slow  and the network topology remains unchanged for long
%time.
%%%\textcolor{red}{and $O_{jk}(t)$ are almost identical to $\delta_{jk}$.} 
%Before topology changes, 
%$\theta_l$ ($l\ge 3$)  decay much faster than $\theta_2$.
%%%Thus, second smallest eigenvalue plays an important role in this case.
%%%We put $\tilde \theta_l^2 \approx \delta_{l2}$, and
%%%$f(\lambda) \langle \tilde \theta^2(\lambda) \rangle \approx
%%%f_2(\lambda) /N$, where $f_2(\lambda)$ is the distribution funcition for
%%%the second smallest eigenvalue (Fig. \ref{fig3}(b)).
%%% 
%Therefore, we can approximate the characteristic time using the average of
%the second smallest eigenvalues as Eq.~(\ref{eq:T_sl})
%%%\begin{align}
%%%%%\frac{\tau_P}{T_{SL}} = \sigma \langle \lambda_{2}\rangle + \mathcal
%%%%% O(\sigma^2).
%%%T_{SL} = -\frac{\tau_P}{\langle \log(1-\sigma \lambda_2)\rangle} .
%%%\label{eq:T_sl}
%%%\end{align}
%%%}
%%%\\

In summary, we have presented a model of interactions between
moving agents that takes into account two different time scales: 
one 
related to local 
synchronization in clusters and the other related to the 
{topology change.}
%speed of the agents. 
%We show that when the second time scale is greater than the first
%more time is required for the system to achieve synchronization
%than when the opposite is true.
We have shown that when the second time scale is greater than the first,
more time is required for the system to achieve synchronization than
the prediction of FSA. %the approximation normally used 
This new effect is particularly important, because
%There exists an intermediate regime where 
it affects the optimal parameter values 
in terms of synchronization time and efficiency.
%This result suggests that the designers of mobile device networks
%should take into account the time scale of agent motion in addition to
%the static topological properties of their systems.}\\
%
%The proposed framework describes evolution of the
%network by allowing agents within a certain distance to
%bring their phases into closer agreement. 
%Our framework correctly predicts 
%the experimentally determined scaling laws of phase decay to the 
%synchronized state for three different regimes. 
%
%Our results are supported by a detailed analysis of the eigenvalues
%of the time dependent Laplacian matrix.
%\cite{ghb08,ygc09}.
Although our model assumes purely random motion, it 
could be easily extended to more realistic patterns of motion \cite{ghb08}.
Our result suggests that 
the interplay between 
instantaneous topology, agent motion,
and interaction rules
plays an important role for the performance of  %network.
mobile systems 
such as ad hoc networks or sensor networks. 
%should take into account %the effect of
% agent motion 
%in addition to the static topological properties of their systems,
%, which 
%since it causes the deviation from FSA.
% the sycnhronization time can be worse due to the deviation from
%FSA close to the percolation transition point.

N.F. and J.K. are grateful for financial support from WGL (project
ECONS) and FET Open project SUMO (grant agreement 266722).
A.D.-G. acknowledges 
financial support from Spanish MCINN (grants FIS2006-13321, FIS2009-13730, PR2008-0114)
and Generalitat de Catalunya (grant 2009SGR00838).

%\footnote{
%\textcolor{red}{I combined some references.}
%}

\bibliographystyle{apsrev}
%\bibliographystyle{aps}
%\bibliography{sync_updated}

\begin{thebibliography}{24}
\expandafter\ifx\csname natexlab\endcsname\relax\def\natexlab#1{#1}\fi
\expandafter\ifx\csname bibnamefont\endcsname\relax
  \def\bibnamefont#1{#1}\fi
\expandafter\ifx\csname bibfnamefont\endcsname\relax
  \def\bibfnamefont#1{#1}\fi
\expandafter\ifx\csname citenamefont\endcsname\relax
  \def\citenamefont#1{#1}\fi
\expandafter\ifx\csname url\endcsname\relax
  \def\url#1{\texttt{#1}}\fi
\expandafter\ifx\csname urlprefix\endcsname\relax\def\urlprefix{URL }\fi
\providecommand{\bibinfo}[2]{#2}
\providecommand{\eprint}[2][]{\url{#2}}

\bibitem[{\citenamefont{Albert and Barab\'asi}(2002)}]{ba02}
\bibinfo{author}{\bibfnamefont{R.}~\bibnamefont{Albert}} \bibnamefont{and}
  \bibinfo{author}{\bibfnamefont{A.-L.} \bibnamefont{Barab\'asi}},
  \bibinfo{journal}{Rev. Mod. Phys.} \textbf{\bibinfo{volume}{74}},
  \bibinfo{pages}{47} (\bibinfo{year}{2002}).

\bibitem[{\citenamefont{{Boccaletti}
  et~al.}(2006{\natexlab{a}})\citenamefont{{Boccaletti}, {Latora}, {Moreno},
  {Chavez}, and {Hwang}}}]{blmch06}
\bibinfo{author}{\bibfnamefont{S.}~\bibnamefont{{Boccaletti}}},
  \bibinfo{author}{\bibfnamefont{V.}~\bibnamefont{{Latora}}},
  \bibinfo{author}{\bibfnamefont{Y.}~\bibnamefont{{Moreno}}},
  \bibinfo{author}{\bibfnamefont{M.}~\bibnamefont{{Chavez}}}, \bibnamefont{and}
  \bibinfo{author}{\bibfnamefont{D.-U.} \bibnamefont{{Hwang}}},
  \bibinfo{journal}{Phys. Rep.} \textbf{\bibinfo{volume}{424}},
  \bibinfo{pages}{175} (\bibinfo{year}{2006}{\natexlab{a}}).

\bibitem[{\citenamefont{Arenas et~al.}(2008)\citenamefont{Arenas,
  D{\'\i}az-Guilera, Kurths, Moreno, and Zhou}}]{adkmz08}
\bibinfo{author}{\bibfnamefont{A.}~\bibnamefont{Arenas}},
  \bibinfo{author}{\bibfnamefont{A.}~\bibnamefont{D{\'\i}az-Guilera}},
  \bibinfo{author}{\bibfnamefont{J.}~\bibnamefont{Kurths}},
  \bibinfo{author}{\bibfnamefont{Y.}~\bibnamefont{Moreno}}, \bibnamefont{and}
  \bibinfo{author}{\bibfnamefont{C.}~\bibnamefont{Zhou}},
  \bibinfo{journal}{Phys. Rep.} \textbf{\bibinfo{volume}{469}},
  \bibinfo{pages}{93} (\bibinfo{year}{2008}).

\bibitem[{\citenamefont{{Sachtjen} et~al.}(2000)\citenamefont{{Sachtjen},
  {Carreras}, and {Lynch}}}]{scl00}
\bibinfo{author}{\bibfnamefont{M.~L.} \bibnamefont{{Sachtjen}}},
  \bibinfo{author}{\bibfnamefont{B.~A.} \bibnamefont{{Carreras}}},
  \bibnamefont{and} \bibinfo{author}{\bibfnamefont{V.~E.}
  \bibnamefont{{Lynch}}}, \bibinfo{journal}{Phys. Rev. E}
  \textbf{\bibinfo{volume}{61}}, \bibinfo{pages}{4877} (\bibinfo{year}{2000}).

\bibitem[{\citenamefont{Olfati-Saber et~al.}(2007)\citenamefont{Olfati-Saber,
  Fax, and Murray}}]{ofm07}
\bibinfo{author}{\bibfnamefont{R.}~\bibnamefont{Olfati-Saber}},
  \bibinfo{author}{\bibfnamefont{J.~A.} \bibnamefont{Fax}}, \bibnamefont{and}
  \bibinfo{author}{\bibfnamefont{R.~M.} \bibnamefont{Murray}},
  \bibinfo{journal}{P. IEEE} \textbf{\bibinfo{volume}{95}},
  \bibinfo{pages}{215} (\bibinfo{year}{2007}).

\bibitem[{\citenamefont{Onnela et~al.}(2007)\citenamefont{Onnela, Saramaki,
  Hyvonen, Szabo, Lazer, Kaski, Kertesz, and Barabasi}}]{onnela07}
\bibinfo{author}{\bibfnamefont{J.-P.} \bibnamefont{Onnela}},
  \bibinfo{author}{\bibfnamefont{J.}~\bibnamefont{Saramaki}},
  \bibinfo{author}{\bibfnamefont{J.}~\bibnamefont{Hyvonen}},
  \bibinfo{author}{\bibfnamefont{G.}~\bibnamefont{Szabo}},
  \bibinfo{author}{\bibfnamefont{D.}~\bibnamefont{Lazer}},
  \bibinfo{author}{\bibfnamefont{K.}~\bibnamefont{Kaski}},
  \bibinfo{author}{\bibfnamefont{J.}~\bibnamefont{Kertesz}}, \bibnamefont{and}
  \bibinfo{author}{\bibfnamefont{A.~L.} \bibnamefont{Barabasi}},
  \bibinfo{journal}{Proc. Natl. Acad. Sci. USA} \textbf{\bibinfo{volume}{104}},
  \bibinfo{pages}{7332} (\bibinfo{year}{2007}).%; 

\bibitem[{\citenamefont{Valencia et~al.}(2008)\citenamefont{Valencia,
  Martinerie, Dupont, and Chavez}}]{vmdc08}
\bibinfo{author}{\bibfnamefont{M.}~\bibnamefont{Valencia}},
  \bibinfo{author}{\bibfnamefont{J.}~\bibnamefont{Martinerie}},
  \bibinfo{author}{\bibfnamefont{S.}~\bibnamefont{Dupont}}, \bibnamefont{and}
  \bibinfo{author}{\bibfnamefont{M.}~\bibnamefont{Chavez}},
  \bibinfo{journal}{Phys. Rev. E} \textbf{\bibinfo{volume}{77}},
  \bibinfo{pages}{050905R} (\bibinfo{year}{2008}).%; %.

\bibitem[{\citenamefont{{Buscarino} et~al.}(2006)\citenamefont{{Buscarino},
  {Fortuna}, {Frasca}, and {Rizzo}}}]{bffr06}
\bibinfo{author}{\bibfnamefont{A.}~\bibnamefont{{Buscarino}}},
  \bibinfo{author}{\bibfnamefont{L.}~\bibnamefont{{Fortuna}}},
  \bibinfo{author}{\bibfnamefont{M.}~\bibnamefont{{Frasca}}}, \bibnamefont{and}
  \bibinfo{author}{\bibfnamefont{A.}~\bibnamefont{{Rizzo}}},
  \bibinfo{journal}{Chaos} \textbf{\bibinfo{volume}{16}},
  \bibinfo{pages}{015116} (\bibinfo{year}{2006}).

%\bibitem[{\citenamefont{Tanner et~al.}(2003)\citenamefont{Tanner, Jadbabaie,
%  and Pappas}}]{tjp03}
%\bibinfo{author}{\bibfnamefont{H.~G.} \bibnamefont{Tanner}},
%  \bibinfo{author}{\bibfnamefont{A.}~\bibnamefont{Jadbabaie}},
%  \bibnamefont{and} \bibinfo{author}{\bibfnamefont{G.~J.}
%  \bibnamefont{Pappas}}, in \emph{\bibinfo{booktitle}{Decision and Control,
%  2003. Proceedings. 42nd IEEE Conference on}} (\bibinfo{year}{2003}),
%  vol.~\bibinfo{volume}{2}, pp. \bibinfo{pages}{2016--2021 Vol.2}.

\bibitem[{\citenamefont{Tanner et~al.}(2003)\citenamefont{Tanner, Jadbabaie,
  and Pappas}}]{tjp03}
\bibinfo{author}{\bibfnamefont{H.~G.} \bibnamefont{Tanner}},
  \bibinfo{author}{\bibfnamefont{A.}~\bibnamefont{Jadbabaie}},
  \bibnamefont{and} \bibinfo{author}{\bibfnamefont{G.~J.}
  \bibnamefont{Pappas}}, in \emph{\bibinfo{booktitle}{Proceedings. 42nd IEEE Conference on Decision and Control,}} 
%  vol.~\bibinfo{volume}{2},
 pp. \bibinfo{pages}{2016--2021} (\bibinfo{year}{2003}).

\bibitem[{\citenamefont{{Buhl} et~al.}(2006)\citenamefont{{Buhl}, {Sumpter},
  {Couzin}, {Hale}, {Despland}, {Miller}, and {Simpson}}}]{bschdms06}
\bibinfo{author}{\bibfnamefont{J.}~\bibnamefont{{Buhl}}},
  \bibinfo{author}{\bibfnamefont{D.~J.~T.} \bibnamefont{{Sumpter}}},
  \bibinfo{author}{\bibfnamefont{I.~D.} \bibnamefont{{Couzin}}},
  \bibinfo{author}{\bibfnamefont{J.~J.} \bibnamefont{{Hale}}},
  \bibinfo{author}{\bibfnamefont{E.}~\bibnamefont{{Despland}}},
  \bibinfo{author}{\bibfnamefont{E.~R.} \bibnamefont{{Miller}}},
  \bibnamefont{and} \bibinfo{author}{\bibfnamefont{S.~J.}
  \bibnamefont{{Simpson}}}, \bibinfo{journal}{Science}
  \textbf{\bibinfo{volume}{312}}, \bibinfo{pages}{1402} (\bibinfo{year}{2006}).

\bibitem[{\citenamefont{Tanaka}(2007)}]{t07}
\bibinfo{author}{\bibfnamefont{D.}~\bibnamefont{Tanaka}},
  \bibinfo{journal}{Phys. Rev. Lett.} \textbf{\bibinfo{volume}{99}},
  \bibinfo{pages}{134103} (\bibinfo{year}{2007}).

\bibitem[{\citenamefont{Romer}(2001)}]{r01}
\bibinfo{author}{\bibfnamefont{K.}~\bibnamefont{R\"{o}mer},
%  \bibinfo{pages}{173} (\bibinfo{year}{2001}).
in \emph{\bibinfo{booktitle}{\bibinfo{journal}{Proceedings of the 2nd ACM international symposium on Mobile ad hoc networking \& computing}},
}} pp. \bibinfo{pages}{173--182} (\bibinfo{year}{2001}).
%}} (\bibinfo{year}{2001}), pp. \bibinfo{pages}{173--182}.

%\bibitem[{\citenamefont{{Mohammadi} et~al.}(2009)\citenamefont{{Mohammadi},
%  {Oskoee}, {Afsharchi}, {Yazdani}, and {Sahimi}}}]{moays09}
%\bibinfo{author}{\bibfnamefont{H.}~\bibnamefont{{Mohammadi}}},
%  \bibinfo{author}{\bibfnamefont{E.~N.} \bibnamefont{{Oskoee}}},
%  \bibinfo{author}{\bibfnamefont{M.}~\bibnamefont{{Afsharchi}}},
%  \bibinfo{author}{\bibfnamefont{N.}~\bibnamefont{{Yazdani}}},
%  \bibnamefont{and} \bibinfo{author}{\bibfnamefont{M.}~\bibnamefont{{Sahimi}}},
%  \bibinfo{journal}{Int. J. Mod. Phys. C} \textbf{\bibinfo{volume}{20}},
%  \bibinfo{pages}{1871} (\bibinfo{year}{2009}).

\bibitem[{\citenamefont{{Sivrikaya} et~al.}(2004)\citenamefont{{Sivrikaya}, and
  {Yener}}}]{sy04}
\bibinfo{author}{\bibfnamefont{F.}~\bibnamefont{{Sivrikaya}}},
  \bibnamefont{and} \bibinfo{author}{\bibfnamefont{B.}~\bibnamefont{{Yener}}},
  \bibinfo{journal}{IEEE Network} \textbf{\bibinfo{volume}{18}},
  \bibinfo{pages}{45} (\bibinfo{year}{2004}).

\bibitem[{\citenamefont{{Belykh} et~al.}(2004)\citenamefont{{Belykh}, {Belykh},
  and {Hasler}}}]{bbh04b}
\bibinfo{author}{\bibfnamefont{I.~V.} \bibnamefont{{Belykh}}},
  \bibinfo{author}{\bibfnamefont{V.~N.} \bibnamefont{{Belykh}}},
  \bibnamefont{and} \bibinfo{author}{\bibfnamefont{M.}~\bibnamefont{{Hasler}}},
  \bibinfo{journal}{Physica D} \textbf{\bibinfo{volume}{195}},
  \bibinfo{pages}{188} (\bibinfo{year}{2004}).

\bibitem[{\citenamefont{Frasca et~al.}(2008)\citenamefont{Frasca, Buscarino,
  Rizzo, Fortuna, and Boccaletti}}]{frasca08}
\bibinfo{author}{\bibfnamefont{M.}~\bibnamefont{Frasca}},
  \bibinfo{author}{\bibfnamefont{A.}~\bibnamefont{Buscarino}},
  \bibinfo{author}{\bibfnamefont{A.}~\bibnamefont{Rizzo}},
  \bibinfo{author}{\bibfnamefont{L.}~\bibnamefont{Fortuna}}, \bibnamefont{and}
  \bibinfo{author}{\bibfnamefont{S.}~\bibnamefont{Boccaletti}},
  \bibinfo{journal}{Phys. Rev. Lett.} \textbf{\bibinfo{volume}{100}},
  \bibinfo{pages}{044102} (\bibinfo{year}{2008}).

\bibitem[{\citenamefont{Porfiri et~al.}(2006)\citenamefont{Porfiri, Stilwell,
  Bollt, and Skufca}}]{psbs06}
\bibinfo{author}{\bibfnamefont{M.}~\bibnamefont{Porfiri}},
  \bibinfo{author}{\bibfnamefont{D.~J.} \bibnamefont{Stilwell}},
  \bibinfo{author}{\bibfnamefont{E.~M.} \bibnamefont{Bollt}}, \bibnamefont{and}
  \bibinfo{author}{\bibfnamefont{J.~D.} \bibnamefont{Skufca}},
  \bibinfo{journal}{Physica D: Nonlinear Phenomena}
  \textbf{\bibinfo{volume}{224}}, \bibinfo{pages}{102 } (\bibinfo{year}{2006}).

\bibitem[{\citenamefont{Stilwell et~al.}(2006)\citenamefont{Stilwell, Bollt,
  and Roberson}}]{sbr06}
\bibinfo{author}{\bibfnamefont{D.~J.} \bibnamefont{Stilwell}},
  \bibinfo{author}{\bibfnamefont{E.~M.} \bibnamefont{Bollt}}, \bibnamefont{and}
  \bibinfo{author}{\bibfnamefont{D.~G.} \bibnamefont{Roberson}},
  \bibinfo{journal}{SIAM J. App. Dyn. Syst.} \textbf{\bibinfo{volume}{5}},
  \bibinfo{pages}{140} (\bibinfo{year}{2006}).

\bibitem[{\citenamefont{Peruani et~al.}(2010)\citenamefont{Peruani, Nicola, and
  Morelli}}]{pnm10}
\bibinfo{author}{\bibfnamefont{F.}~\bibnamefont{Peruani}},
  \bibinfo{author}{\bibfnamefont{E.~M.} \bibnamefont{Nicola}},
  \bibnamefont{and} \bibinfo{author}{\bibfnamefont{L.~G.}
  \bibnamefont{Morelli}}, \bibinfo{journal}{New J. Phys.}
  \textbf{\bibinfo{volume}{12}}, \bibinfo{pages}{093029}
  (\bibinfo{year}{2010}).

\bibitem[{\citenamefont{Kuramoto}(1984)}]{Kuramoto84}
\bibinfo{author}{\bibfnamefont{Y.}~\bibnamefont{Kuramoto}},
  \emph{\bibinfo{title}{Chemical oscillations, waves, and turbulence}}
  (\bibinfo{publisher}{Springer-Verlag}, \bibinfo{address}{New York, NY, USA},
  \bibinfo{year}{1984}); %.
%
%\bibitem[{\citenamefont{{Acebr{\'o}n} et~al.}(2005)\citenamefont{{Acebr{\'o}n},
%  {Bonilla}, {P{\'e}rez-Vicente}, {Ritort}, and {Spigler}}}]{abprs05}
\bibinfo{author}{\bibfnamefont{J.~A.} \bibnamefont{{Acebr{\'o}n}}},
  \bibinfo{author}{\bibfnamefont{L.~L.} \bibnamefont{{Bonilla}}},
  \bibinfo{author}{\bibfnamefont{C.~J.} \bibnamefont{{P{\'e}rez-Vicente}}},
  \bibinfo{author}{\bibfnamefont{F.}~\bibnamefont{{Ritort}}}, \bibnamefont{and}
  \bibinfo{author}{\bibfnamefont{R.}~\bibnamefont{{Spigler}}},
  \bibinfo{journal}{Rev. Mod. Phys.} \textbf{\bibinfo{volume}{77}},
  \bibinfo{pages}{137} (\bibinfo{year}{2005}).

\bibitem[{\citenamefont{{Filatrella} et~al.}(2008)\citenamefont{{Filatrella},
  {Nielsen}, and {Pedersen}}}]{fnp07}
\bibinfo{author}{\bibfnamefont{G.}~\bibnamefont{{Filatrella}}},
  \bibinfo{author}{\bibfnamefont{A.~H.} \bibnamefont{{Nielsen}}},
  \bibnamefont{and} \bibinfo{author}{\bibfnamefont{N.~F.}
  \bibnamefont{{Pedersen}}}, \bibinfo{journal}{Europ. Phys. J. B}
  \textbf{\bibinfo{volume}{61}}, \bibinfo{pages}{485} (\bibinfo{year}{2008}); %.
%
%\bibitem[{\citenamefont{{L{\"a}mmer} et~al.}(2006)\citenamefont{{L{\"a}mmer},
%  {Kori}, {Peters}, and {Helbing}}}]{lkph06}
\bibinfo{author}{\bibfnamefont{S.}~\bibnamefont{{L{\"a}mmer}}},
  \bibinfo{author}{\bibfnamefont{H.}~\bibnamefont{{Kori}}},
  \bibinfo{author}{\bibfnamefont{K.}~\bibnamefont{{Peters}}}, \bibnamefont{and}
  \bibinfo{author}{\bibfnamefont{D.}~\bibnamefont{{Helbing}}},
  \bibinfo{journal}{Physica A} \textbf{\bibinfo{volume}{363}},
  \bibinfo{pages}{39} (\bibinfo{year}{2006}); %.
%
%\bibitem[{\citenamefont{Klein et~al.}(2008)\citenamefont{Klein, Lee, Morgansen,
%  and Javidi}}]{klmj08}
\bibinfo{author}{\bibfnamefont{D.~J.} \bibnamefont{Klein}},
  \bibinfo{author}{\bibfnamefont{P.}~\bibnamefont{Lee}},
  \bibinfo{author}{\bibfnamefont{K.~A.} \bibnamefont{Morgansen}},
  \bibnamefont{and} \bibinfo{author}{\bibfnamefont{T.}~\bibnamefont{Javidi}},
  \bibinfo{journal}{IEEE Journal on Selected Areas in Communications}
  \textbf{\bibinfo{volume}{26}}, \bibinfo{pages}{695} (\bibinfo{year}{2008}).

\bibitem[{\citenamefont{Dall and Christensen}(2002)}]{dc02}
\bibinfo{author}{\bibfnamefont{J.}~\bibnamefont{Dall}} \bibnamefont{and}
  \bibinfo{author}{\bibfnamefont{M.}~\bibnamefont{Christensen}},
  \bibinfo{journal}{Phys. Rev. E} \textbf{\bibinfo{volume}{66}},
  \bibinfo{pages}{016121} (\bibinfo{year}{2002}); %.
%
%\bibitem[{\citenamefont{Balister et~al.}(2005)\citenamefont{Balister,
%  Bollob\'{a}s, and Walters}}]{bbw05}
\bibinfo{author}{\bibfnamefont{P.}~\bibnamefont{Balister}},
  \bibinfo{author}{\bibfnamefont{B.}~\bibnamefont{Bollob\'{a}s}},
  \bibnamefont{and} \bibinfo{author}{\bibfnamefont{M.}~\bibnamefont{Walters}},
  \bibinfo{journal}{Random Struct. Algor.} \textbf{\bibinfo{volume}{26}},
  \bibinfo{pages}{392} (\bibinfo{year}{2005}).

\bibitem{anime}
See supplementary material for animations.


\bibitem[{\citenamefont{{Boccaletti}
  et~al.}(2006{\natexlab{b}})\citenamefont{{Boccaletti}, {Hwang}, {Ch{\'a}vez},
  {Amann}, {Kurths}, and {Pecora}}}]{bhcakp06}
\bibinfo{author}{\bibfnamefont{S.}~\bibnamefont{{Boccaletti}}},
  \bibinfo{author}{\bibfnamefont{D.-U.} \bibnamefont{{Hwang}}},
  \bibinfo{author}{\bibfnamefont{M.}~\bibnamefont{{Ch{\'a}vez}}},
  \bibinfo{author}{\bibfnamefont{A.}~\bibnamefont{{Amann}}},
  \bibinfo{author}{\bibfnamefont{J.}~\bibnamefont{{Kurths}}}, \bibnamefont{and}
  \bibinfo{author}{\bibfnamefont{L.~M.} \bibnamefont{{Pecora}}},
  \bibinfo{journal}{Phys. Rev. E} \textbf{\bibinfo{volume}{74}},
  \bibinfo{pages}{016102} (\bibinfo{year}{2006}{\natexlab{b}});
%\bibitem[
%{\citenamefont{Zhao et~al.}(2009)\citenamefont{Zhao, Hill, and
%  Liu}}]{zhl09}
\bibinfo{author}{\bibfnamefont{J.}~\bibnamefont{Zhao}},
  \bibinfo{author}{\bibfnamefont{D.~J.} \bibnamefont{Hill}}, \bibnamefont{and}
  \bibinfo{author}{\bibfnamefont{T.}~\bibnamefont{Liu}},
  \bibinfo{journal}{Automatica} \textbf{\bibinfo{volume}{45}},
  \bibinfo{pages}{2502 } (\bibinfo{year}{2009}).

\bibitem[{\citenamefont{Gonzalez et~al.}(2008)\citenamefont{Gonzalez, Hidalgo,
  and Barabasi}}]{ghb08}
\bibinfo{author}{\bibfnamefont{M.~C.} \bibnamefont{Gonzalez}},
  \bibinfo{author}{\bibfnamefont{C.~A.} \bibnamefont{Hidalgo}},
  \bibnamefont{and} \bibinfo{author}{\bibfnamefont{A.-L.}
  \bibnamefont{Barabasi}}, \bibinfo{journal}{Nature}
  \textbf{\bibinfo{volume}{453}}, \bibinfo{pages}{779} (\bibinfo{year}{2008}).

\end{thebibliography}

\end{document}